\documentclass[12pt]{article}
\usepackage{a4,amsmath,theorem,epsfig}%,times,euler,euscript}
\usepackage{multirow}
\renewcommand{\baselinestretch}{1.1}
\newcommand{\Appendix}[1]{Appendix~\ref{#1}}   
\theoremstyle{break}\theorembodyfont{\rmfamily}
                    \newtheorem{Alg}{Algorithm}[subsection]
\newcommand{\Algorithm}[1]{Algorithm \ref{#1}}   

\newcommand{\Table}[1]{Table~\ref{#1}}

\newcommand{\ben}{\begin{enumerate}}
\newcommand{\een}{\end{enumerate}}
\newcommand{\bit}{\begin{itemize}}
\newcommand{\eit}{\end{itemize}}
\newcommand{\bc}{\begin{center}}
\newcommand{\ec}{\end{center}}
\newcommand{\bq}{\begin{equation}}
\newcommand{\eq}{\end{equation}}
\newcommand{\bqa}{\begin{eqnarray}}
\newcommand{\eqa}{\end{eqnarray}}

\newcommand{\nl}{\notag\\}

\def\demo{$\varDelta\eta\hspace{-1pt}\mu\acute{o}\kappa\hspace{-1pt}%
           \varrho\iota\tau\hspace{-1pt}o\hspace{0.5pt}\varsigma$}

\newcommand{\df}{:=}
\newcommand{\fd}{=:}
\newcommand{\Ant}{A}

\newcommand{\vhi}{\varphi}
\newcommand{\Ss}{s}
\newcommand{\sa}{\sigma}
\newcommand{\Sa}{\varSigma}

\newcommand{\la}{\lambda}
\newcommand{\ve}{\varepsilon}
\newcommand{\vecp}{\hspace{-1pt}\vec{\hspace{1pt}p}}
\newcommand{\vecq}{\hspace{-1pt}\vec{\hspace{1pt}q}}
\newcommand{\vecv}{\hspace{-1pt}\vec{\hspace{1pt}v}}

\newcommand{\si}{\mathrm{s}}
\newcommand{\co}{\mathrm{c}}

\newcommand{\srac}[2]{{\textstyle\frac{#1}{#2}}}
\newcommand{\Pih}{\Pi^{(-1/2)}}
\newcommand{\ao}{a_1}
\newcommand{\aoz}{\ao^{(0)}}
\newcommand{\aop}{\ao^{+}}
\newcommand{\aom}{\ao^{-}}
\newcommand{\aopm}{\ao^{\pm}}
\newcommand{\aot}{\tilde{a}_1}
\newcommand{\fao}{f}
\newcommand{\at}{a_2}
\newcommand{\atb}{\bar{a}_2}
\newcommand{\atz}{\at^{(0)}}
\newcommand{\atp}{\at^{+}}
\newcommand{\atm}{\at^{-}}
\newcommand{\atpm}{\at^{\pm}}
\newcommand{\atbp}{\atb^{+}}
\newcommand{\atbm}{\atb^{-}}
\newcommand{\atbpm}{\atb^{\pm}}

\newcommand{\hh}{h}
\newcommand{\ho}{h_1}
\newcommand{\vv}{v}
\newcommand{\ww}{w}
\newcommand{\lar}{\leftarrow}
\begin{document}
\begin{center}
%
%%%%%%%%%%%%%%%%%%%%%%%%%%%%%%%%% title %%%%%%%%%%%%%%%%%%%%%%%%%%%%%%%%%%%%%%%
{\bf\Huge A hierarchical phase space generator}\\\vspace{0.25\baselineskip}
{\bf\Huge for QCD antenna structures}

\vspace{2\baselineskip}%%%%%%%%% authors %%%%%%%%%%%%%%%%%%%%%%%%%%%%%%%%%%%%%%
{\Large Andr\'e van Hameren%
        \footnote{E-mail: {\tt andrevh@inp.demokritos.gr}} 
    and Costas G.~Papadopoulos%
        \footnote{E-mail: {\tt Costas.Papadopoulos@cern.ch}}
}

\vspace{0.25\baselineskip}%%%%%% adress %%%%%%%%%%%%%%%%%%%%%%%%%%%%%%%%%%%%%%%
{\it\large Institute of Nuclear Physics, NCSR \demo, 15310 Athens, Greece}

\vspace{0.25\baselineskip}%%%%%%% date %%%%%%%%%%%%%%%%%%%%%%%%%%%%%%%%%%%%%%%%
{\large\today}

\renewcommand{\baselinestretch}{1}
\vspace{2\baselineskip}%%%%%%%% abstract %%%%%%%%%%%%%%%%%%%%%%%%%%%%%%%%%%%%%%
{\bf Abstract}\\\vspace{0.5\baselineskip}
\parbox{0.8\linewidth}{\small\hspace{15pt}%
We present a
``hierarchical'' strategy for phase space generation in
order to efficiently map the antenna momentum structures, 
typically occurring in  QCD amplitudes.
}
\end{center}
\vspace{\baselineskip}

%%%%%%%%%%%%%%%%%%%%%%%%%%%%%%%%%%%%%%%%%%%%%%%%%%%%%%%%%%%%%%%%%%%%%%%%%%%%%%%
\section{Introduction}
%%%%%%%%%%%%%%%%%%%%%%%%%%%%%%%%%%%%%%%%%%%%%%%%%%%%%%%%%%%%%%%%%%%%%%%%%%%%%%%
The reliable description of multi-jet production at the LHC~\cite{lhc} will be
an important issue. This is not only related to the study of QCD in multi-parton
final states but it is also very important in order to estimate several
backgrounds for new physics effects. For instance, new unstable massive
particles that decay to many partons may be discovered at the LHC only when a
reliable description of these final states is established.

In this respect, apart from the problem of computing scattering matrix elements
with many particles, also the efficient phase space generation is of great
importance, because the scattering amplitudes in QCD exhibit strong peaking
structures in phase space, which have to be taken into account by the
generation algorithm. Flat phase space generators, like {\tt
RAMBO}~\cite{rambo}, will not be adequate for this task. In the last years
several methods to efficiently integrate the peaking structures of the
scattering amplitudes have emerged, and have been used in several contexts~\cite{LEP2}. For
instance, {\tt PHEGAS}~\cite{phegas} is an example where an efficient,
automated, mapping of all possible peaking structures of a given scattering
process has been established. The algorithm is based on the ``natural''
mappings dictated by the Feynman graphs contributing to the given process, so
that the number of kinematical channels used to generate the phase space
is equal to the number of Feynman graphs. Using adaptive methods, like
multi-channel optimization~\cite{KleissPittau} and by throwing away channels
that are negligible, we may end up with a few channel generator exhibiting high
efficiency, as is indeed the case in $n(+\gamma)$-fermion production in
$e^+e^-$ collisions. In contrast, the QCD scattering amplitudes point towards
the opposite direction: large number of Feynman graphs which means large number
of kinematical channels which, moreover, contribute equally to the result.

A way out off this problem may be based on the long-standing remark that
$n+2$-gluon amplitude may be described by a very compact expression when
special helicities are assigned to the gluons, which, combined with the leading
color approximation, results to 
\begin{equation}
\sum_{c}|\mathcal{M}|^2=8 \left(\frac{N_{\mathrm{c}}}{2}\right)^{n} (N_{\mathrm{c}}^2-1)
\sum_{1\le i<j}^{n+2} (p_i\cdot p_j)^4 
\sum_{P(2,\ldots,n+2)}\Ant_{n+2}(p_1,\ldots,p_{n+2})
\;\;,
\end{equation}
where $N_{\mathrm{c}}$ refers to the number of colors, 
\begin{equation}
\Ant_{n+2}(p_1,\ldots,p_{n+2})
  \df
[\;(p_1\cdot p_2)(p_2\cdot p_3)
         \cdots(p_{n+1}\cdot p_{n+2})(p_{n+2}\cdot p_1)\;]^{-1}
%\frac{1}{(p_1\cdot p_2)(p_2\cdot p_3)
%         \cdots(p_{n+1}\cdot p_{n+2})(p_{n+2}\cdot p_1)}
\;\;,
\label{defAnt}\end{equation}
and the sum over all permutations of the $2^{\textrm{nd}}$ to the 
$(n+2)^{\textrm{nd}}$ argument of this function is taken, with the exception 
of those
that are equivalent under reflection $i\mapsto n+4-i$ \cite{kuijf}. 

{\tt SARGE} \cite{sarge}
is the first known example of a phase space generator that deals
with the momentum structures entering the above expression, namely with 
(\ref{defAnt}),
known as {\it antenna structures}. The algorithm is based on the
``democratic'' strategy to generate the $n$ body phase space, as is the case
for {\tt RAMBO}, and it makes use of the scale symmetry of the antenna
to achieve the required goal.

In this paper, we study the ``hierarchical'' strategy for phase space
generation in order to efficiently map the momentum antenna structures.  The
idea is as follows. Using the standard two-body phase space (neglecting factors of $2\pi$) 
\begin{equation}
d\Phi_2(P;s_1,s_2;p_1,p_2)
  \df 
d^4p_1\,\delta_+(p_1^2-s_1)\,d^4p_2\,\delta_+(p_2^2-s_2)\,\delta^4(P-p_1-p_2)
\;\;,
\label{twobody}\end{equation}
we decompose the phase space as
\begin{align}
d\Phi_n(P;p_1\ldots,p_n) 
  \df&\,
\Big(\prod_{i=1}^n d^4p_i\,\delta_+(p_i^2-\sa_i)\Big)
\delta^4\Big(\sum_{i=1}^{n} p_i-P\Big)
\notag\\
  =&\, 
d\Ss_{n-1}\,d\Phi_2(Q_n;\sa_{n},\Ss_{n-1};p_n,Q_{n-1}) \notag\\
  \times&\,
d\Ss_{n-2}\,d\Phi_2(Q_{n-1};\sa_{n-1},\Ss_{n-2};p_{n-1},Q_{n-2})\notag\\
  &\hspace{20pt}\vdots\notag\\
  \times&\, 
d\Ss_2\,d\Phi_2(Q_3;\sa_3,\Ss_2;p_3,Q_2)\;d\Phi_2(Q_2;\sa_2,\sa_1;p_2,p_1)
\;.
\label{psdec}\end{align}
%where an overall factor $(2\pi)^{4-3n}$ is assumed.  
The task is to express the
phase space in terms of the invariants $p_i\cdot p_j$ appearing in the antenna
structure (\ref{defAnt}), so that, using a suitable mapping, we can construct a
density that, apart from constant and soft terms, will be identical to this
antenna structure.

In the first section, we describe the basic building block of the algorithm,
which is the expression of the two-body phase space in terms of the scaled invariants.
In the second section, we demonstrate how this basic building block can be used
in a sequential way to produce the full antenna. Finally in the third section,
some details concerning the numerical implementation of the algorithm as well
as comparisons with known generators is given. The Appendices present all
relevant generation algorithms from which the exact functioning of the
generator can be reconstructed. 

%%%%%%%%%%%%%%%%%%%%%%%%%%%%%%%%%%%%%%%%%%%%%%%%%%%%%%%%%%%%%%%%%%%%%%%%%%%%%%%
\section{The hierarchical antenna}
%%%%%%%%%%%%%%%%%%%%%%%%%%%%%%%%%%%%%%%%%%%%%%%%%%%%%%%%%%%%%%%%%%%%%%%%%%%%%%%
\subsection{The basic building block\label{SecBBB}}
%%%%%%%%%%%%%%%%%%%%%%%%%%%%%%%%%%%%%%%%%%%%%%%%%%%%%%%%%%%%%%%%%%%%%%%%%%%%%%%
To illustrate the idea we consider the generation of the 
2-body phase space (\ref{twobody})
when two massless antenna momenta $q_1,q_2$ are given. 
The momentum $P$ can be decomposed as
\begin{equation}
P^\mu=r q_1^\mu + A_1^\mu\fd A_1^\mu +B_1^\mu\;\;\;
\mathrm{with}\;\;\; r=P^2/(2 P\cdot q_1)  \;\;,
\notag
\end{equation}
so that the Sudakov parameterization of $p_1$ is given by
\begin{equation}
p_{1}^\mu=a_{1} A_{1}^\mu + b_{1} B_{1}^\mu + k_{1}^\mu  \;\;.
\notag
\end{equation}
where the variables $a_1$ and $b_1$ are given by 
\begin{equation}
a_1=\frac{p_1\cdot B_1}{A_1\cdot B_1}\quad,\quad
b_1=\frac{p_1\cdot A_1}{A_1\cdot B_1} \;\;.
\notag
\end{equation}
The same can be done in terms of $p_2$ and $q_2$, and in the center-of-mass
frame (CMF) of $P$, where $P=(\sqrt{s}\,,0,0,0)$, 
$\cos(\angle(\vecq_1,\vecq_2))=\co$ and $\si=\sqrt{1-\co^2}$, one can choose 
\begin{alignat}{2}
   A_1&=\srac{1}{2}\sqrt{s}(1,0,0,-1) 
       \;\;&,\quad  A_2&=\srac{1}{2}\sqrt{s}(1,0,-\si,-\co)\;\;,   \notag\\
   B_1&=\srac{1}{2}\sqrt{s}(1,0,0,1)     
       \;\;&,\quad  B_2&=\srac{1}{2}\sqrt{s}(1,0,\si,\co)\;\;,      \notag\\
   k_1&=(0,x_1,y_1,0)  
       \;\;&,\quad  k_2&=(0,x_2,y_2\co,-y_2\si)\;\;.   \notag
\end{alignat}
The phase space can now be completely expressed in
terms of $a_1$ and $a_2$, leading to
\begin{equation} 
d\Phi_2(P;p_1,p_2)=
da_1\,da_2\, \Pih\Theta(\Pi)  \;\;,
\label{2bps}
\end{equation}
with
\begin{align}
\Pi(a_1,a_2) &= 
   4\si^2[(1-a_2+\bar{s}_2-\bar{s}_1)a_2-\bar{s}_2] \notag\\
    &-[(1-2a_1-\bar{s}_1+\bar{s}_2)+(1-2a_2-\bar{s}_1+\bar{s}_2)\co]^2 \;\;,
\label{defPi}\end{align}
where $\bar{s}_{1,2}=s_{1,2}/s$,
% \begin{multline}
% \Pi=4 \si^2 \left( a_2\left(1-a_2+\frac{s_2-s_1}{s}\right)
% -\frac{s_2}{s}\right) \\
% -\left( \left(1-2 a_1-\frac{s_1-s_2}{s}\right)
% +\left(1-2 a_2+\frac{s_2-s_1}{s}\right) \co\right)^2
% \label{Pi}
% \end{multline}
and where $\Theta$ is the step function.
%\[ b_1=1-a_1+\frac{m_1^2-m_2^2}{4 Q^2} \]
%\[ b_2=1-a_2+\frac{m_2^2-m_1^2}{4 Q^2} \]
%and of course
In terms of Lorentz invariants, the parameters $a_1$ and $a_2$ are given by
\begin{equation}
a_1=\frac{q_1\cdot p_1}{q_1\cdot P}\quad,\quad
a_2=\frac{p_2\cdot q_2}{P\cdot q_2} \;\;.
\label{a1a2}
\end{equation}
So in order to obtain a two-body phase space with a density which depends
on the invariants $a_1,a_2$ following some given function $f(a_1,a_2)$, one 
has to generate $a_1,a_2$ following a density proportional to 
$f(a_1,a_2)\times\Pih(a_1,a_2)$ in the region where $\Pi(a_1,a_2)>0$, and 
construct the momenta following the Sudakov parameterization. Explicitly, 
the direct construction is given by
\begin{align}
   p_1^0 &\lar (s + s_1 - s_2)/(2\sqrt{s})        
%   \quad,&\quad 
%   p_2^0 &= \sqrt{s}-p_1^0                       
   \quad,\notag\\
   p_1^3 &\lar p_1^0 - \sqrt{s}\,a_1                      
%   \quad,&\quad
%   p_2^3 &= -p_1^3                               
   \quad,\notag\\
   p_1^2 &\lar (\;(\sqrt{s}-p_1^0-\sqrt{s}\,a_2)+\co p_1^3\;)/\si  
%   \quad,&\quad
%   p_2^2 &= - p_1^2                           
   \quad,\notag\\
   p_1^1 &\lar \epsilon(\;(p_1^0)^2-s_1-(p_1^2)^2-(p_1^3)^2\;)^{1/2}
%   \quad,&\quad
%   p_2^1 &= -p_1^1                               
   \quad,\notag
\end{align}
where $\epsilon$ should be a fair random variable which can take values $+1$
and $-1$.  For more details about this procedure, we refer the reader to the
\Appendix{AppA}.

%%%%%%%%%%%%%%%%%%%%%%%%%%%%%%%%%%%%%%%%%%%%%%%%%%%%%%%%%%%%%%%%%%%%%%%%%%%%%%%
\subsection{Antenna generation}
%%%%%%%%%%%%%%%%%%%%%%%%%%%%%%%%%%%%%%%%%%%%%%%%%%%%%%%%%%%%%%%%%%%%%%%%%%%%%%%
\newcommand{\gf}{g}
\newcommand{\Gf}{G}
\newcommand{\Bb}{B}
\newcommand{\Dd}{D}
\newcommand{\Ln}{L}
\newcommand{\Ff}{F}
In the hierarchical/sequential approach, the generation strategy proceeds
through a sequence of two-body phase space generations following the
decomposition (\ref{psdec}). At each two-body generation, one final-state
momentum $p_{k}$ is generated, together with the sum $Q_{k-1}$ of the remaining
final-state momenta to be generated. This suggest to label the momenta in a way
opposite to the order of generation, so first $p_{n},Q_{n-1}$ are generated,
then $p_{n-1},Q_{n-2}$ and so on.  
The starting point is
the CMF of the initial momenta $q_1$ and $q_2$ with $Q_n=q_1+q_2$ being the
overall momentum. The CMF of momentum $Q_k$ we denote by CMF$_k$. The pair
$p_{k},Q_{k-1}$ is generated by generating variables $\ao^{(k)},\at^{(k)}$
and constructing the momenta as described before.
These variables are now equal to
\begin{equation}
\ao^{(k)}=\frac{p_{k+1}\cdot p_k}{p_{k+1}\cdot Q_k}
\quad
\mathrm{and}
\quad
\at^{(k)}=\frac{q_2\cdot Q_{k-1}}{q_2\cdot Q_k} \;\;.
\notag\end{equation}
This happens in CMF$_k$, so in 
order to obtain $p_{k},Q_{k-1}$, the constructed momenta have to be boosted 
such that $(\sqrt{Q_k^2},0,0,0)$ is transformed to $Q_k$.

We would like to generate the momenta following a
density that is proportional to 
\begin{equation}
\Ant_{n+2}(q_1,p_n\ldots,p_1,q_2)=
[\;(q_1\cdot p_n)\,(p_n\cdot p_{n-1})\,\ldots (p_1\cdot q_2)\;]^{-1}
%\frac{1}{(q_1\cdot p_n)\,(p_n\cdot p_{n-1})\,\ldots (p_1\cdot q_2)} 
\;\;.
\label{int}
\end{equation}
Since the integrand is infra-red singular, a cutoff on the invariants is
necessary. Therefore, we define a symmetric matrix $\sigma_{ij}$ which encodes 
the restrictions on the momenta through
\begin{equation}
 \sigma_{ii} =\sigma_i=p_i^2 \quad,\quad
 \sqrt{\sigma_i \sigma_j}\le \sigma_{ij}\le p_i\cdot p_j 
\quad\textrm{and}\quad
   \Sa_k\df\sum_{i=1}^k \sigma_i 
\;\;.
\label{sigma}
\end{equation}
% We shall also need sums of the diagonal elements, which we denote by 
% \begin{equation}
%    \Sa_k=\sum_{i=1}^k \sigma_i \;\;.
%b \notag\end{equation}
Before we proceed, we do three observations. Firstly, we have
\begin{equation}
  p_{k+1}\cdot Q_k=(Q^2_{k+1}-Q^2_{k}-p^2_{k+1})/2
\;\;. 
\notag\end{equation}
Secondly, we have
\begin{equation}
  \frac{\Ss_{k+1}-\Sa_{k+1}}
       {\left(\Ss_{k+1}-\sa_{k+1}-s_{k}\right)\left(\Ss_{k}-\Sa_{k}\right)}
  = \frac{d}{d\Ss_k}\,
    \log\left(\frac{\Ss_{k}-\Sa_{k}}{\Ss_{k+1}-\sa_{k+1}-\Ss_{k}}\right)
  \fd \gf_{k+1}(\Ss_k) 
  \;\;,
\label{s1g}
\end{equation}
and thirdly, we can write
\begin{equation}
\Ant_{n+2}(q_1,p_n\ldots,p_1,q_2)=
\frac{1}{2^{n-1}} 
(\Ss_n-\Sa_n)(q_1\cdot Q_n)(q_2\cdot Q_n)
\left(\prod_{k=n}^{3}\gf_{k}(\Ss_{k-1})
\frac{1}{\ao^{(k)}\at^{(k)}}\right)\frac{1}{\ao^{(2)}\at^{(2)}}  \;\;,
\notag
\end{equation}
with $\Ss_{n}=Q^2_{n}$, $p_{n+1}=q_1$ and $Q_1=p_1$.
These observations suggest that
the phase space generation 
\begin{alignat}{3}
&
d\Ss_{n-1}\, \gf_{n}(\Ss_{n-1}) \;\;
&&
da^{(n)}_1\, \frac{1}{a^{(n)}_1}\;\; 
&&
da^{(n)}_2\, \frac{1}{a^{(n)}_2}\;\; 
\Pih_{(n)} \Theta (\Pi_{(n)}) 
\nl
&
d\Ss_{n-2}\, \gf_{n-1}(\Ss_{n-2})\;\; 
&&
da^{(n-1)}_1\, \frac{1}{a^{(n-1)}_1}\;\; 
&&
da^{(n-1)}_2\, \frac{1}{a^{(n-1)}_2}\;\; 
\Pih_{(n-1)} \Theta (\Pi_{(n-1)})
\nl
&&
\vdots 
&&&
\nl 
&
d\Ss_2\, \gf_{3}(\Ss_2)\;\;
&&
da^{(3)}_1 \, \frac{1}{a^{(3)}_1}\;\;
&&
da^{(3)}_2 \, \frac{1}{a^{(3)}_2}\;\; 
\Pih_{(3)} \Theta (\Pi_{(3)})
\nl 
&&
&
da^{(2)}_1 \, \frac{1}{a^{(2)}_1}\;\;
&&
da^{(2)}_2 \, \frac{1}{a^{(2)}_2}\;\; 
\Pih_{(2)} \Theta (\Pi_{(2)}) \;\;,
\label{closedA}
\end{alignat}
will lead to a density for the momenta that is proportional to $\Ant_{n+2}$.
Three variables are generated in each CMF$_k$, namely $\Ss_{k-1}$, $a^{(k)}_1$
and $a^{(k)}_2$.  
Just as the integration of $\Ss_{k-1}$ (\ref{s1g}), 
also the integration of $\ao^{(k)},\at^{(k)}$ 
results 
in a volume factor
that depends on the corresponding variables generated in CMF$_{k+1}$.  As we
will show in \Appendix{AppA}, however, these factors are logarithmic functions
of their arguments and exhibit a non-singular behavior, and we call them {\em
soft\/} factors. The total actual density will therefore be the product of $n-1$
soft factors times the antenna structure under consideration. 

In the end, we want to generate all permutations in the momenta of (\ref{int}).
Those for which $q_1$ and $q_2$ each appear in two factors (none of which is
$q_1\cdot q_2$) cannot be obtained by simple re-labeling. In order to obtain
these, we observe that they can be decomposed
into two antennas, namely
\begin{equation}
        \Ant_{m+2}(q_1,p_{m},p_{m-1},\ldots,p_{2},p_{1},q_2)
  \times\Ant_{n-m+2}(q_2,p_{n},p_{n-1},\ldots,p_{m+2},p_{m+1},q_1) 
%        \Ant_{m+2}(q_1,p_{1},p_{2},\ldots,p_{m-1},p_{m},q_2)
%  \times\Ant_{n-m+2}(q_1,p_{m+1},p_{m+2},\ldots,p_{n-1},p_{n},q_2) 
\label{otherA}
\end{equation}
and each of these can be generated after the 
decomposition,
\begin{align}{}
d\Phi_n(P;p_1\ldots,p_n)&=
d\Ss_m\, d\Ss_{n-m}\, 
d\Phi_2(Q_n;\Ss_{m},\Ss_{n-m};Q_m,Q_{n-m})
\nl
&\times
d\Phi_m(Q_m;p_1,\ldots,p_m)\,
d\Phi_{n-m}(Q_{n-m}
;p_{m+1},\ldots,p_n)  \;\;.
\label{split}
\end{align}
In order to combine the two sub-antennas to the required antenna structure, we
have to take into account in the first decomposition a density that is
proportional to 
\begin{equation}
%\begin{alignat}{1}
%&d\Ss_m\, dQ_{n-m}^2\, 
%d\Phi_2(Q_n;Q_m,Q_{n-m}) {\cal D'} =d\Ss_m\, d\Ss_{n-m}\,d\cos\theta\, d\phi\,
%\nl
%&
% \frac{\lambda^{1/2}(\Ss,\Ss_m,\Ss_{n-m})}{\Ss}
\frac{\Theta(\,\sqrt{\Ss_{n}}-\sqrt{\Ss_m}-\sqrt{\Ss_{n-m}}\;)}{
(q_1\cdot Q_m)
(q_1\cdot Q_{n-m})
(q_2\cdot Q_m)
(q_2\cdot Q_{n-m})
\,
\Ss_m
\,
\Ss_{n-m}} \;\;.
\label{splitden}
%\end{alignat}
\end{equation}
The case of $m=1$ is still special. Then, the first step in 
(\ref{closedA}) should be replaced by
\begin{equation}
d\Ss_{n-1}g'(\Ss_{n-1}) \;\;
da^{(n)}_1\, \;\; 
da^{(n)}_2\, \;\; 
\Pih_{(n)} \Theta (\Pi_{(n)})\;
\frac{1}{a^{(n)}_1\,a^{(n)}_2\,(1-a^{(n)}_1)(1-a^{(n)}_2)} \;\;,
\label{1split}
\end{equation}
and the rest of the sequence should go on  with the replacement of $p_n$ by
$p_{n+1}$ at the second step. For the density $g'$ we refer to \Appendix{AppA}. 
There, we have collected all integrals and generation
algorithms related to the antenna generation described so far.

%%%%%%%%%%%%%%%%%%%%%%%%%%%%%%%%%%%%%%%%%%%%%%%%%%%%%%%%%%%%%%%%%%%%%%%%%%%%%%%
\subsection{Open antennas}
%%%%%%%%%%%%%%%%%%%%%%%%%%%%%%%%%%%%%%%%%%%%%%%%%%%%%%%%%%%%%%%%%%%%%%%%%%%%%%%
As it will be clear from the numerical analysis presented in the next section,
the soft factors appearing in the description of the QCD antenna contribute to
a certain extent to the variance of the Monte Carlo integration. There is an
alternative approach, that still follow the  hierarchical/sequential generation strategy, 
and give better results. It based on the observation that the production
of an `open' antenna structure, namely one where the last product $q_2\cdot
p_1$ is missing, is simpler, since it can be constructed without using the
variables $a^{(k)}_2$.  They can be replaced by flatly generated azimuthal
angles, to that $da^{(k)}_2 \Pih_{(k)} \Theta (\Pi_{(k)})\rightarrow
d\vhi^{(k)}$.  The basic decomposition therefore becomes
\begin{multline}
        d\Ss_{n-1}\gf_{n}(\Ss_{n-1})\, 
	    d\vhi^{(n)}\,da^{(n)}_1\frac{1}{\ao^{(n)}}\times 
        d\Ss_{n-2}\gf_{n-1}(\Ss_{n-2})\, 
	    d\vhi^{(n-1)}\,da^{(n-1)}_1\frac{1}{\ao^{(n-1)}} 
	 \times\cdots \\ \cdots\times
        d\Ss_{2}\gf_{3}(\Ss_{2})\, 
	    d\vhi^{(3)}\,da^{(3)}_1\frac{1}{\ao^{(3)}}\times
	  d\vhi^{(2)}  \,da^{(2)}_1\frac{1}{\ao^{(2)}}   \;\;.
\label{openA}\end{multline}
% \begin{alignat}{2}
%         &d\Ss_{n-1}\gf_{n}(\Ss_{n-1}) 
% 	    &&d\vhi^{(n)}\,da^{(n)}_1\frac{1}{a_1^{(n)}} \nl
%         &d\Ss_{n-2}\gf_{n-1}(\Ss_{n-2}) 
% 	    &&d\vhi^{(n-1)}\,da^{(n-1)}_1\frac{1}{a_1^{(n-1)}} \nl
% 	& &\vdots&                                     \nl
%         &d\Ss_{2}\gf_{3}(\Ss_{2}) 
% 	    &&d\vhi^{(3)}\,da^{(3)}_1\frac{1}{a_1^{(3)}}\nl
% 	&  &&d\vhi^{(2)}  \,da^{(2)}_1\frac{1}{a_1^{(2)}}   \;\;.
% \label{openA}\end{alignat}
% In case all momenta are massless, the functions $\gf_{k}$ are the same as 
% before. The massive case is described in \Appendix{AppA}.
% In that general case, the final density becomes proportional to 
% \begin{equation}
%  \Dd_0(Q_n;p_n,\ldots,p_1) \sim 
%   \frac{(q_{1}\cdot Q_n)\sqrt{\lambda(\Ss_n,\sa_n,\Sa_{n-1})}}
% 	 {(q_{1}\cdot p_{n})(p_{n}\cdot p_{n-1})\cdots
%           (p_{3}\cdot p_{2})(p_{2}\cdot p_{1})} 
%     \quad,
% \label{openA0}\end{equation}
% with the K\"ahlen function 
% \begin{equation}
% \lambda(x,y,z)=x^2+y^2+z^2-2xy-2xz-2yz \;\;,
% \label{kahlen}
% \end{equation}
% so that the term proportional to $p_1\cdot q_2$ is obviously missing, which is
% the reason why we call this an `open' antenna. 
This way, we will get the antenna density (\ref{int}) without the factor 
$p_1\cdot q_2$ in 
the denominator, which is the reason why we call this an `open' antenna.
A `closed' antenna can be
obtained using the fact that, by combining two open antennas, one can choose
for another factor from the antenna string to be missing.  Then, a
multi-channeling procedure can be performed with these different choices,
leading to a density that is, roughly speaking, proportional to 
\begin{equation}
   \frac{(q_{1}\cdot p_{n})+(p_{n}\cdot p_{n-1})+\cdots
         +(p_{3}\cdot p_{2})+(p_{2}\cdot p_{1})+(p_1\cdot q_2)}
   {(q_{1}\cdot p_{n})(p_{n}\cdot p_{n-1})\cdots
          (p_{3}\cdot p_{2})(p_{2}\cdot p_{1})(p_1\cdot q_2)}   \;\;.
\notag\end{equation}
To get the different choices, a first splitting of $Q_{n}$ into $Q_{m}$ and
$Q_{n-m}$ has to be performed, after which open antennas are generated from
each of these, one with $q_1$ and the other with $q_2$ as initial momentum. 
% The momenta $Q_{m}$ and $Q_{n-m}$ should be generated with a density such that
% the factors coming from the numerator of (\ref{openA0}) in the generation 
% of the antennas are canceled.
For details, we refer to \Appendix{AppB}.

%%%%%%%%%%%%%%%%%%%%%%%%%%%%%%%%%%%%%%%%%%%%%%%%%%%%%%%%%%%%%%%%%%%%%%%%%%%%%%%
\section{Results}
%%%%%%%%%%%%%%%%%%%%%%%%%%%%%%%%%%%%%%%%%%%%%%%%%%%%%%%%%%%%%%%%%%%%%%%%%%%%%%%
\newcommand{\e}[2]{$#1\times10^{#2}$}
In this section, we present results obtained by {\tt SARGE} and {\tt
HAAG}\footnote{{\tt HAAG} stands for: {\tt Hierarchical AntennA
Generation}.}, the program that implements the hierarchical algorithm
described before. In order to be as general as possible, the only cut we
apply is 
\begin{equation}
   (p_i+p_j)^2 \ge s_0\;\;,
\notag\end{equation}
where $i,j(i\neq j)$ runs from $1$ to $n+2$ where $n$ is the number of
final-state particles. Unless explicitly mentioned differently, we use
$s_0=900\,\mathrm{GeV}^2$ and the total energy
$\sqrt{s}=1000\,\mathrm{GeV}$.  Moreover, all particles are assumed to be
massless in order to compare with {\tt SARGE}, with which only massless
particles can be treated. 

As it was mentioned in the introduction, we are interested in integrating sums
of QCD antenna structures (\ref{defAnt}). We start by considering the simplest
case, namely integrating the function 
\begin{equation}
 s^2 
 [\;(p_1\cdot p_3)(p_3\cdot p_4)(p_4\cdot p_2)(p_2\cdot p_5)
        \ldots(p_{n+2}\cdot p_1)\;]^{-1}
\label{ss3}
\end{equation}
that corresponds to a given permutation of the momenta, namely
$(1,3,4,2,5,\ldots,n+2)$.  In \Table{ss3t} we give the results for {\tt SARGE},
{\tt HAAG} with open antenna generation, and {\tt HAAG(C)} with closed antenna
generation. In all three codes the same single channel, corresponding to
(\ref{ss3}), has been used in the generation. $N_{\mathrm{gen}}$ and
$N_{\mathrm{acc}}$ are the number of generated and accepted events, and by $f$
we define 
\begin{equation}
  f\df \frac{V_2}{I^2} \;\;,
\notag\end{equation}
where $V_2$ is the quadratic
variance and $I$ is the estimated integral. $f$ is clearly a measure of the
efficiency of the generator. Moreover $\ve$, defined as 
\begin{equation}
  \varepsilon\df\frac{<w>}{w_{\mathrm{max}}} \;\;,
\notag\end{equation}
is the usual generation efficiency related
for instance to `unweighted' events in a realistic simulation.
%-------------------------------------------------------------------------------
\begin{table}[ht]
\begin{center}
\begin{tabular}{|c|c|c|c|c|c|c|c|}
\hline
jets & algorithm  & $N_{\mathrm{gen}}$  &  $N_{\mathrm{acc}}$  &  $I$         &   $\Delta I$  & $f$  &          $\varepsilon$(\%)  \\ \hline 
\multirow{3}{1cm}{\centering 4}
    & {\tt SARGE}   &  \e{1}{5}  & 34853   & \e{.251}{-9} & \e{.734}{-11} &     $85.9 $        & $0.34$ \\
    & {\tt HAAG}    &  \e{5}{4}  & 31193   & \e{.260}{-9} & \e{.280}{-11} &     $5.75 $        & $1.77$   \\
    & {\tt HAAG(C)} &  \e{5}{4}  & 28366   & \e{.256}{-9} & \e{.252}{-11} &     $4.84 $        & $4.22$   \\\hline
\multirow{3}{1cm}{\centering 5}
    & {\tt SARGE}   &  \e{2.5}{5}  & 30960   & \e{.438}{-10} & \e{.153}{-11} &     $307  $        & $0.23$ \\
    & {\tt HAAG}    &  \e{6.5}{4}  & 29855   & \e{.442}{-10} & \e{.640}{-12} &     $13.6 $        & $1.02$   \\
    & {\tt HAAG(C)} &  \e{6.5}{4}  & 24345   & \e{.441}{-10} & \e{.706}{-12} &     $16.7 $        & $1.04$   \\\hline
\multirow{3}{1cm}{\centering 6}
    & {\tt SARGE}   &  \e{1}{6}    & 28383   & \e{.487}{-11} & \e{.164}{-12} &     $1141 $        & $0.21$ \\
    & {\tt HAAG}    &  \e{1.2}{5}  & 32070   & \e{.487}{-11} & \e{.658}{-13} &     $21.9 $        & $1.48$   \\
    & {\tt HAAG(C)} &  \e{1.2}{5}  & 25040   & \e{.485}{-11} & \e{.886}{-13} &     $40.1 $        & $0.69$   \\\hline
\end{tabular}
\caption[.]{Results for the single-channel integration/generation.}
\label{ss3t} 
\vspace{-15pt}
\end{center}
\end{table}
%-------------------------------------------------------------------------------
The results agree well, and exhibit the fact that the generated densities of
the generators the hierarchical type are much closer to the integrand.
Moreover, the closed antenna algorithm {\tt HAAG(C)} becomes less efficient
compared to the open one as the number of particles increases. The same
picture is reproduced for an arbitrary permutation.

For a realistic QCD calculation, the integrated function may be approximated
by a sum over permutations. Therefore, an efficient generator has to include
all possible channels, where each channel corresponds to a given permutation
of the momenta.
% Moreover a multi-channeling optimization is indispensable in that case. Of
% course {\tt HAAG} incorporates such a multi-channeling procedure. 
In that case, a multi-channeling optimization procedure can be applied, which
is incorporated in {\tt HAAG}. In order to study the efficiency of this
optimization we consider the same integration as before, but with all channels
contributing to the generation and allowing for optimization. In this
optimization procedure, we discard channels that contribute less than a certain
pre-determined fraction to the set of available channels. It is expected, of
course, that in end the right permutation will be `chosen' by the optimization.
This is indeed the case and the results are presented in \Table{as3}. We see
that the optimization results to a picture close to the one obtained with the
single channel generation, with some noticeable improvement in the case of {\tt
SARGE}. We also include results with {\tt SARGE.n}, a slightly different
version, described in \Appendix{AppSARGE}.
%-------------------------------------------------------------------------------
\begin{table}[ht]
\begin{center}
\begin{tabular}{|c|c|c|c|c|c|c|c|}
\hline
jets & algorithm  & $N_{\mathrm{gen}}$  &  $N_{\mathrm{acc}}$  &  $I$         &   $\Delta I$  & $f$  &          $\varepsilon$(\%)  \\ \hline 
\multirow{3}{1cm}{\centering 4}
    & {\tt SARGE}   &  \e{1}{5}  & 52516   & \e{.262}{-9} & \e{.294}{-11} &     $12.6 $         & $1.29$ \\
  & {\tt SARGE.n}   &  \e{1}{5}  & 46529   & \e{.260}{-9} & \e{.298}{-11} &     $13.2 $         & $1.55$ \\
    & {\tt HAAG}    &  \e{5}{4}  & 34293   & \e{.257}{-9} & \e{.210}{-11} &     $3.36 $         & $4.28$   \\
    & {\tt HAAG(C)} &  \e{5}{4}  & 29736   & \e{.259}{-9} & \e{.227}{-11} &     $3.84 $         & $3.91$   \\\hline
\multirow{3}{1cm}{\centering 5}
    & {\tt SARGE}   &  \e{2.5}{5}  & 32315   & \e{.422}{-10} & \e{.106}{-11} &     $159  $        & $0.44$ \\
  & {\tt SARGE.n}   &  \e{2. }{5}  & 30994   & \e{.440}{-10} & \e{.807}{-12} &     $67.2 $        & $0.83$ \\
    & {\tt HAAG}    &  \e{6.5}{4}  & 31063   & \e{.444}{-10} & \e{.503}{-12} &     $8.32 $        & $1.17$   \\
    & {\tt HAAG(C)} &  \e{6.5}{4}  & 24179   & \e{.436}{-10} & \e{.593}{-12} &     $12.03$        & $1.84$   \\\hline
\multirow{3}{1cm}{\centering 6}
    & {\tt SARGE}   &  \e{1}{6}    & 29138   & \e{.476}{-11} & \e{.145}{-12} &     $933 $       & $0.45$   \\
  & {\tt SARGE.n}   &  \e{1}{6}    & 35445   & \e{.492}{-11} & \e{.109}{-12} &     $492$        & $0.25$   \\
    & {\tt HAAG}    &  \e{1.2}{5}  & 33278   & \e{.483}{-11} & \e{.595}{-13} &     $18.2 $      & $1.19$   \\
    & {\tt HAAG(C)} &  \e{1.2}{5}  & 24126   & \e{.471}{-11} & \e{.749}{-13} &     $30.3 $      & $1.21$   \\\hline
\end{tabular} 
\caption[.]{Results for the all-channel generation with optimization.}
\label{as3} 
\vspace{-15pt}
\end{center}
\end{table}
%-------------------------------------------------------------------------------

As is the case for any multi-channel generator, a computational complexity
problem arises when the number of channels increases. For instance, in our case
we are facing a  number of $\frac{1}{2}(n+1)!$ channels! On the other hand, it
is also clear that the channels we are considering have a large overlap in most
of the available phase space. It is therefore worth to investigate the
dependence of the integration efficiency on the number of channels used. This
is presented in \Table{cha}, where the full antenna
\begin{equation}
 s^2 
 \sum_{P(2,\ldots,n+2)}
 [\;(p_1\cdot p_3)(p_3\cdot p_4)(p_4\cdot p_2)(p_2\cdot p_5) 
        \ldots(p_{n+2}\cdot p_1)\;]^{-1}  
\label{aa}
\end{equation}
is integrated, using a number of channels that has been selected on a
random basis. We see the rather interesting phenomenon
that a decent description can be achieved with a much smaller number of
channels.    
%-------------------------------------------------------------------------------
\begin{table}[ht]
\begin{center}
\begin{tabular}{|c||c|c|c|c|c|c|c|}
\hline
\# channels & $2520$     & $1500$  &  $1000$  &  $500$     &  $200$    &     $50$   &  $10$    \\ \hline 
 $f$    & $5.33$     & $5.37$  &  $5.48$  &  $5.72$    &  $6.14$    &    $11.6$  &  $84.7$  \\ \hline
 $N_{\mathrm{acc}}$  & $26630$    & $26521$ &  $26437$ &  $26676$   &  $27009$   &    $27190$ &  $27205$ \\ \hline
 $\varepsilon(\%)$  
        & $11.2$     & $13.1$  &  $11.6$  &  $7.1$     &  $7.5$     &    $1.7$   &  $0.28$  \\ \hline
\end{tabular} 
\caption[.]{All-channel integration with subsets of channels for generation.}
\label{cha} 
\vspace{-15pt}
\end{center}
\end{table}
%-------------------------------------------------------------------------------
Variations of this technique of using only subsets of channels, for example 
choosing another subset after each step of multi-channel optimization, lead to 
the same picture.

The complete results of the integration of the full antenna 
are presented in \Table{aa30}.
%-------------------------------------------------------------------------------
\begin{table}[ht]
\begin{center}
\begin{tabular}{|c|c|c|c|c|c|c|c|}
\hline
jets & algorithm  & $N_{\mathrm{gen}}$  &  $N_{\mathrm{acc}}$  &  $I$         &   $\Delta I$  & $\varepsilon$(\%)& $f$    \\ \hline 
\multirow{2}{1cm}{\centering 4}
    & {\tt SARGE}&  \e{1}{5}  & 47483   & \e{.166}{-7} & \e{.115}{-9} &     $4.21 $        & $4.8$ \\
    & {\tt HAAG} &  \e{6}{4}  & 42019   & \e{.167}{-7} & \e{.810}{-10} &    $12.01$        & $1.4$   \\\hline
\multirow{2}{1cm}{\centering 5}
    & {\tt SARGE}&  \e{3}{5}    & 39095   & \e{.176}{-7} & \e{.162}{-9} &     $3.27 $        & $25.6$ \\
    & {\tt HAAG} &  \e{1.2}{5}  & 55234   & \e{.177}{-7} & \e{.856}{-10} &    $7.53 $        & $2.7$   \\\hline
\multirow{2}{1cm}{\centering 6}
    & {\tt SARGE}&  \e{1.5}{6}  & 44529   & \e{.157}{-7} & \e{.135}{-9} &     $2.95 $        & $109$ \\
    & {\tt HAAG} &  \e{1.8}{5}  & 47911   & \e{.161}{-7} & \e{.905}{-10} &    $7.15 $        & $5.7$   \\\hline
\multirow{2}{1cm}{\centering 7}
    & {\tt SARGE}&  \e{1}{7}    & 47766   & \e{.123}{-7} & \e{.988}{-10} &     $3.02 $        & $642$ \\
    & {\tt HAAG} &  \e{3.6}{5}  & 45599   & \e{.123}{-7} & \e{.241}{-10} &     $5.11 $        & $13$   \\\hline
\multirow{2}{1cm}{\centering 8}
    & {\tt SARGE}&  \e{1}{8}    & 53560   & \e{.784}{-8} & \e{.554}{-10} &    $3.29$        & $4998$ \\
    & {\tt HAAG} &  \e{1}{6}   & 49206   & \e{.789}{-8} & \e{.496}{-10} &    $6.30$        & $39$   \\\hline
\end{tabular} 
\caption[.]{Results for the all-channel integration.}
\label{aa30}
\vspace{-15pt}
\end{center}
\end{table}
%-------------------------------------------------------------------------------
We see that {\tt HAAG} has a much
better $f$ factor than {\tt SARGE}. On the other hand the $\ve$ exhibits a less
dramatic effect. This is related to the fact that {\tt SARGE} generates a phase
space that is much larger than the one defined by the cut on $s_0$. In that
sense, if the main time consumption in a given computation is spent over the
evaluation of the integrand (matrix element squared), it is more fair to
compare the square of the estimated expected error, normalized by the number of
accepted events $N_{\mathrm{acc}}$. In that case we see that {\tt HAAG} is still 2-3 times
more efficient, and if we consider a smaller cut, namely $\sqrt{s_0}=10$ GeV,
this gain goes up to an order of magnitude (\Table{aa10}). 
%-------------------------------------------------------------------------------
\begin{table}[ht]
\begin{center}
\begin{tabular}{|c|c|c|c|c|c|c|c|}
\hline
jets & algorithm  & $N_{\mathrm{gen}}$  &  $N_{\mathrm{acc}}$  &  $I$         &   $\Delta I$  & $\varepsilon$(\%)& $f$    \\ \hline 
\multirow{2}{1cm}{\centering 4}
    & {\tt SARGE}&  \e{1}{5}  & 60986   & \e{.364}{-6} & \e{.548}{-8} &     $0.631 $        & $22.7$ \\
    & {\tt HAAG} &  \e{6}{4}  & 46763   & \e{.366}{-6} & \e{.235}{-8} &     $4.34$          & $2.47$   \\\hline
\multirow{2}{1cm}{\centering 5}
    & {\tt SARGE}&  \e{2}{5}    & 43150   & \e{.619}{-6} & \e{.165}{-7} &     $0.29 $        & $142$ \\
    & {\tt HAAG} &  \e{1}{5}    & 56034   & \e{.643}{-6} & \e{.465}{-8} &    $1.84 $         & $5.23$   \\\hline
\multirow{2}{1cm}{\centering 6}
    & {\tt SARGE}&    \e{1}{6}  & 67811   & \e{.114}{-5} & \e{.257}{-7} &     $0.28 $        & $502$ \\
    & {\tt HAAG} &  \e{1.4}{5}  & 51983   & \e{.111}{-5} & \e{.883}{-8} &     $2.50 $        & $8.83$   \\\hline
\multirow{2}{1cm}{\centering 7}
    & {\tt SARGE}&  \e{5}{6}    & 84391   & \e{.186}{-5} & \e{.346}{-7} &     $0.176 $        & $1723$ \\
    & {\tt HAAG} &  \e{2}{5}    & 44015   & \e{.192}{-5} & \e{.177}{-7} &     $2.24 $        & $16$   \\\hline
\multirow{2}{1cm}{\centering 8}
    & {\tt SARGE}&  \e{5}{7}    & 175541   & \e{.354}{-5} & \e{.517}{-7} &    $.119$        & $10618$ \\
    & {\tt HAAG} &  \e{5}{5}    & 58874    & \e{.350}{-5} & \e{.289}{-7} &    $1.65$        & $34$   \\\hline
\end{tabular} 
\caption[.]%
        {Results for the all-channel integration%
         with $s_0=100\,\mathrm{GeV}^2$.}
\label{aa10}
\vspace{-10pt}
\end{center}
\end{table}
%-------------------------------------------------------------------------------

For a realistic calculation of the cross section of a QCD process, 
one may assume that the time it takes to perform one
evaluation of the integrand is much larger than the time it takes to generate 
one accepted event and to calculate the weight. This means that the computing 
time is completely determined by the number of accepted events 
$N_{\mathrm{acc}}$. We introduce 
\begin{equation}
   \frac{N_{\mathrm{acc}}f}{N_{\mathrm{gen}}}
\notag\end{equation}
as a measure of the computing time. For a realistic calculation, one has to 
multiply this number by the evaluation time of the integrand, and devide 
by the square of the relative error one wants to reach. Figure 1 shows this 
quantity as 
function of the number of produced partons using the data of \Table{aa10}.
According to this graph, a calculation with {\tt SARGE} would take $10$ times
longer than the calculation with {\tt HAAG}.
%-------------------------------------------------------------------------------
\begin{center}
\epsfig{file=fig.1,width=0.4\linewidth,angle=270}
\end{center}
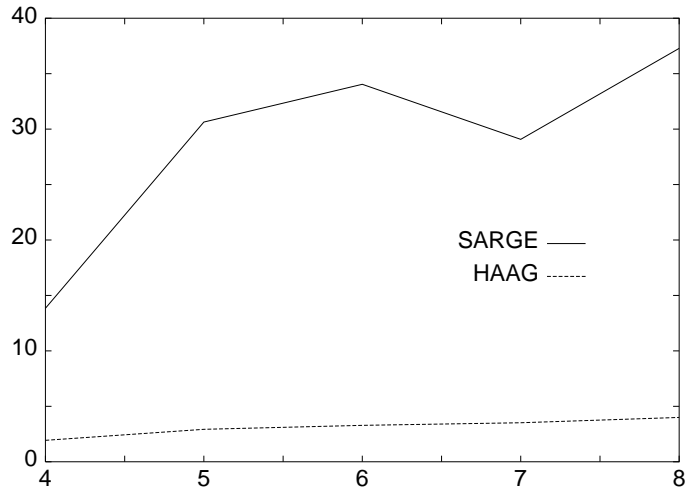
Figure 1: $N_{\mathrm{acc}}f/N_{\mathrm{gen}}$ (a measure of computing time) 
as function of the number of produced partons.\\[2em]
%-------------------------------------------------------------------------------

Finally, Figure 2 shows the dependence of the result on the value of the
infrared cut-off $\sqrt{s_0}$ and the number of produced partons.  
The function
clearly exhibits a negative power behavior for $s_0$. Moreover, the curve
becomes steeper with increasing number of jets, suggesting that at least
the leading power is related to the final-state multiplicity. 
%-------------------------------------------------------------------------------
\begin{center}
\epsfig{file=fig.2,height=8.0cm,width=8.0cm}
\end{center}
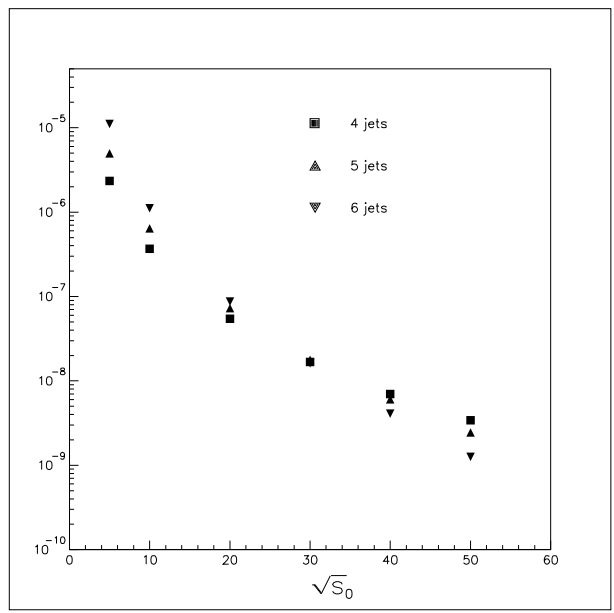
Figure 2: The integral of the full antenna as function of $\sqrt{s_0}$ for 
different values of the number of produced partons.
%-------------------------------------------------------------------------------

%%%%%%%%%%%%%%%%%%%%%%%%%%%%%%%%%%%%%%%%%%%%%%%%%%%%%%%%%%%%%%%%%%%%%%%%%%%%%%%
\section{Conclusions}
%%%%%%%%%%%%%%%%%%%%%%%%%%%%%%%%%%%%%%%%%%%%%%%%%%%%%%%%%%%%%%%%%%%%%%%%%%%%%%%
{\tt HAAG} exhibits an improved efficiency compared to {\tt
SARGE} for multi-parton calculations. It is also more powerful in describing
densities where a partial symmetrization over the permutation space is
considered. Finally, {\tt HAAG} makes no fundamental distinction among massless
and massive particles, so it can be used for an arbitrary multi-partonic
process.

%%%%%%%%%%%%%%%%%%%%%%%%%%%%%%%%%%%%%%%%%%%%%%%%%%%%%%%%%%%%%%%%%%%%%%%%%%%%%%%
\section*{Acknowledgments}
%%%%%%%%%%%%%%%%%%%%%%%%%%%%%%%%%%%%%%%%%%%%%%%%%%%%%%%%%%%%%%%%%%%%%%%%%%%%%%%
The authors would like to thank Ronald Kleiss for proof reading this paper.
The research has been financially supported by the European Union under
contract number HPRN-CT-2000-00149.

%%%%%%%%%%%%%%%%%%%%%%%%%%%%%%%%%%%%%%%%%%%%%%%%%%%%%%%%%%%%%%%%%%%%%%%%%%%%%%%
%   references 
\def\EPJ#1#2#3{Eur.\ Phys.\ J.\ {\bf #1} (#2) #3}
\def\CPC#1#2#3{Comp.\ Phys.\ Comm.\ {\bf #1} (#2) #3}
\def\PR#1#2#3{Phys.\ Rev.\ {\bf{#1}} (#2) #3}
\def\PRL#1#2#3{Phys.\ Rev.\ Lett. {\bf{#1}} (#2) #3}
\def\PL#1#2#3{Phys.\ Lett.\ {\bf{#1}} (#2) #3}
\def\PRep#1#2#3{Phys.\ Rep.\ {\bf{#1}} (#2) #3}
\def\NP#1#2#3{Nucl.\ Phys.\ {\bf{#1}} (#2) #3}
\def\ZP#1#2#3{Z.\ f.\ Phys.\ {\bf{#1}} (#2) #3}
\def\IJMP#1#2#3{Int.\ J.\ Mod.\ Phys.\ {\bf{#1}} (#2) #3}
\def\MPL#1#2#3{Mod.\ Phys.\ Lett.\ {\bf{#1}} (#2) #3}
\def\ibid#1#2#3{{\it ibid} {\bf{#1}} (#2) #3}
%
 
%
%%%%%%%%%%%%%%%%%%%%%%%%%%%%%%%%%%%%%%%%%%%%%%%%%%%%%%%%%%%%%%%%%%%%%%%%%%%%%%%

%%%%%%%%%%%%%%%%%%%%%%%%%%%%%%%%%%%%%%%%%%%%%%%%%%%%%%%%%%%%%%%%%%%%%%%%%%%%%%%
\section*{Appendices}
\renewcommand{\thesubsection}{\Alph{subsection}}
\renewcommand{\thesubsubsection}{\thesubsection\arabic{subsubsection}}
\renewcommand{\theparagraph}{\thesubsubsection.\arabic{paragraph}}
\newcommand{\countpar}{\refstepcounter{paragraph}}
\newcommand{\parcount}{\theparagraph\hspace{10pt}}
%%%%%%%%%%%%%%%%%%%%%%%%%%%%%%%%%%%%%%%%%%%%%%%%%%%%%%%%%%%%%%%%%%%%%%%%%%%%%%%
The following appendices contain details about the generation of the various
random variables necessary to build phase space with the desired density. The
techniques used to achieve this are {\em inversion}, {\em rejection} and {\em
multi-channeling}. For details about these techniques, we refer to
\cite{thesisandre}. We only want to mention that inversion is applicable if one
has an analytic expression (with reasonable complexity) of the inverse of the
indefinite integral of the density. Rejection can be applied if one knows a 
function which is strictly larger than the density and
is proportional to a density one {\em is} able to generate. 
The efficiency is
given by one divided by the integral of that function. Multi-channeling can be
used if the density can be written as the weighted sum of densities, each of
which one {\em is} able to generate, and if the weights are positive.

%%%%%%%%%%%%%%%%%%%%%%%%%%%%%%%%%%%%%%%%%%%%%%%%%%%%%%%%%%%%%%%%%%%%%%%%%%%%%%%
\subsection{Closed antenna\label{AppA}}
%%%%%%%%%%%%%%%%%%%%%%%%%%%%%%%%%%%%%%%%%%%%%%%%%%%%%%%%%%%%%%%%%%%%%%%%%%%%%%%
In this appendix we present all relevant algorithms for the generation of the
closed antenna.  More specifically, in the following we describe the generation
of the three variables $s_{k-1}$, $a^{(k)}_1$ and  $a^{(k)}_2$ that
are needed to describe the antenna at the $k-$th CMF. Most of the time the
superscript $(k)$ is omitted for convenience.

%%%%%%%%%%%%%%%%%%%%%%%%%%%%%%%%%%%%%%%%%%%%%%%%%%%%%%%%%%%%%%%%%%%%%%%%
\subsubsection{Generation of {\boldmath $s_{k-1}$}} 
%%%%%%%%%%%%%%%%%%%%%%%%%%%%%%%%%%%%%%%%%%%%%%%%%%%%%%%%%%%%%%%%%%%%%%%
Since the indefinite integral is given in (\ref{s1g}), we see that 
$\Ss_{k-1}$ can be generated with inversion.
The limits are given by 
\begin{equation}
\Ss_k-\Delta_{k-1}\ge \Ss_{k-1}\ge \Lambda_{k-1} \quad\textrm{with}\quad
\Lambda_{k} = \Sa_k+\sum_{i\neq j}^k\sa_{ij} \;\;,\;\;\;
\Delta_{k} = \sa_{k+1}+2\sum_{i=1}^k\sa_{k+1,i} \;\;,
\label{defsklim}\end{equation}
and the weight factor is
\begin{equation}
%\mathcal{W} =
   \log\left(\frac{\Ss_k-\Sa_{k-1}-\Delta_{k-1}}{\Delta_{k-1}-\sa_k}\right) 
 + \log\left(\frac{\Ss_k-\sa_{k}-\Lambda_{k-1}}
                  {\Lambda_{k-1}-\Sa_{k-1}}\right) \;\;.
\notag\end{equation}
% In case $\sa_k=0$ we may also have 
% \begin{equation}
% \Delta_{k-1}=\max\left(  \Delta_{k-1} ; \frac{\sa_{k+1,k}}{p_{k+1}\cdot Q_k} \Ss_k
% \right)
% \notag\end{equation}
% Moreover in the case of $m=1$ we get simply
In the case of $m=1$ (\ref{1split}) we simply have 
\begin{equation}
\int d\Ss_{k-1} \, \frac{1}{\Ss_{k-1}-\Sa_{k-1}}
 = \log(\Ss_{k-1}-\Sa_{k-1})  \;\;,
\label{gprime}
\end{equation}
so that the weight factor is
\begin{equation}
%\mathcal{W} =
  \log\left( \Ss_k-\Delta_{k-1}-\Sa_{k-1} \right)
-\log\left( \Lambda_{k-1}-\Sa_{k-1} \right) \;\;.
\notag\end{equation}

%%%%%%%%%%%%%%%%%%%%%%%%%%%%%%%%%%%%%%%%%%%%%%%%%%%%%%%%%%%%%%%%%%%%%%%%
\subsubsection{Generation of {\boldmath $\ao,\at$}}
%%%%%%%%%%%%%%%%%%%%%%%%%%%%%%%%%%%%%%%%%%%%%%%%%%%%%%%%%%%%%%%%%%%%%%%%
The general integral corresponding to the generation of $\ao,\at$ is 
given by 
\begin{equation}
%   \mathJ = 
   \int d\ao d\at\,\frac{\Theta(\,\Pi(\ao,\at)\,)\,
                         \Theta(\,\ao-\aoz\,)\,\Theta(\,\at-\atz\,)}
                        {\ao\at\sqrt{\Pi(\ao,\at)}} \;\;,
\label{defJ}\end{equation}
where $\Pi$ is defined in (\ref{defPi}), and $\aoz,\atz$ are 
possibly necessary infra-red cut-offs. We shall analyze this integral by first
integrating over the $\at$-variable, and then over the $\ao$-variable. The 
generation has then to be performed in the opposite order: first $\ao$ and 
then $\at$. As we shall see, the inclusion of the cut-off on $\ao$ does not 
lead to complications, but the inclusion of {\em both\/} cut-offs {\em does\/}.
For that case, we see two solutions. Firstly, we can replace the integral by 
\begin{equation}
%   \mathJ = 
   \int d\ao d\at\,\frac{\Theta(\,\Pi(\ao,\at)\,)\,
                         \Theta(\,\ao-\aoz\,)}
                        {\ao(\at+\hh)\sqrt{\Pi(\ao,\at)}} \;\;,
\notag\end{equation}
where $\hh$ is related to $\atz$. This, of course, changes the actual density
with which $\ao$ and $\at$ are generated, but may, for small $\hh$, still be
considered suitable for the desired antenna structure. Also, the final antenna
generator will cover phase space less efficiently (will generate `to much'
phase space), simply because less cuts are included analytically. In the second
solution, we write the integrand as the sum 
\begin{equation}
%   \mathJ = 
   \frac{\Theta(\,\Pi(\ao,\at)\,)\,\Theta(\,\ao-\aoz\,)}
               {\ao(\at+\ao)\sqrt{\Pi(\ao,\at)}}
	 +\frac{\Theta(\,\Pi(\ao,\at)\,)\,\Theta(\,\at-\atz\,)}
               {(\ao+\at)\at\sqrt{\Pi(\ao,\at)}} \;\;,
\notag\end{equation}
which is exactly equal to the original integrand on the phase space for which
{\em both\/} cuts are included. Both integrands can be integrated analytically,
so that the multi-channeling procedure can be applied to generate their sum.
For this solution, the only problem is that, again, `to much' phase space is
generated.

%%%%%%%%%%%%%%%%%%%%%%%%%%%%%%%%%%%%%%%%%%%%%%%%%%%%%%%%%%%%%%%%%%%%%%%%
\countpar\paragraph{\parcount The {\boldmath $\at$}-variable}
%%%%%%%%%%%%%%%%%%%%%%%%%%%%%%%%%%%%%%%%%%%%%%%%%%%%%%%%%%%%%%%%%%%%%%%%
In order to integrate
over the $\at$-variable, $\Pi(\ao,\at)$ is more conveniently written as
\begin{equation}
  \Pi(\ao,\at) = 4(\,\atp(\ao)-\at\,)(\,\at-\atm(\ao)\,) \;\;,
\end{equation}
with
\begin{multline}
\atpm(\ao)
\df\frac{1}{2}\Big(1+\frac{\Ss_{k-1}-\sa_k}{\Ss_k}
                  +\Big(1-2\ao-\frac{\Ss_{k-1}-\sa_k}{\Ss_k}\Big)\co\Big) \\
 \pm\si\Big(\ao\Big(1-\ao-\frac{\Ss_{k-1}-\sa_k}{s_k}\Big)
                      -\frac{\sa_k}{\Ss_k}\Big)^{1/2}
 \;\;.
\label{defa2pm}\end{multline}
Then, the general $\at$-integral is given by
\begin{equation}
  \int_{\atm}^{\atp}d\at\,\frac{1}{(\at+\hh)\sqrt{(\atp-\at)(\at-\atm)}} 
  \;\;,
\notag\end{equation}
with $\hh=0$ if no cut-off on $\at$ is desirable, and $\hh$ related to $\atz$ or
$\hh=\ao$, depending on the solution mentioned above if a cut-off {\em is}
desirable.
By substituting $\at\leftarrow\at-\hh$, this integral can be written as
\begin{equation}
  \int_{\atbm}^{\atbp}d\at\,\frac{1}{\at\sqrt{(\atbp-\at)(\at-\atbm)}}
  = \frac{-2}{\sqrt{\atbp\atbm}}
    \left[ \arctan\left(\frac{\atbm(\atbp-\at)}{\atbp(\at-\atbm)}\right)
    \right]_{\atbm}^{\atbp}
  = \frac{\pi}{\sqrt{\atbp\atbm}}  
  \;\;,
\notag\end{equation}
with $\atbpm=\atpm+\hh$. The explicit indefinite integral shows that the 
variable $\at$ can be generated by inversion.

%%%%%%%%%%%%%%%%%%%%%%%%%%%%%%%%%%%%%%%%%%%%%%%%%%%%%%%%%%%%%%%%%%%%%%%%
\countpar\paragraph{\parcount The {\boldmath $\ao$}-variable}
%%%%%%%%%%%%%%%%%%%%%%%%%%%%%%%%%%%%%%%%%%%%%%%%%%%%%%%%%%%%%%%%%%%%%%%%
We start this section by mentioning that, in the case 
$p_{k+1}^2=\sigma_{k+1}\not = 0$,
the variables
\begin{equation}
  \ao^{(k)}=\frac{p_{k}\cdot p_{k+1}}{Q_{k}\cdot p_{k+1}}
\notag\end{equation}
can be expressed in terms of the `massless' $\aot^{(k)}$
defined in terms of the `long' component 
$p^{(L)}_{k+1}$ of $p_{k+1}$ in the CMF$_k$: if  
$p^{(L)}_{k+1} = (p_{k+1}^0, \beta_k^{-1}\vecp_{k+1})$ and 
$p^{(S)}_{k+1} = (p_{k+1}^0,-\beta_k^{-1}\vecp_{k+1})$ with 
\begin{equation}
% p^{(L)}_{k+1} = \Big(p_{k+1}^0,\frac{\vecp_{k+1}}{\beta_k}\Big)
% \;,\;\;
% p^{(S)}_{k+1} = \Big(p_{k+1}^0,-\frac{\vecp_{k+1}}{\beta_k}\Big) 
% \quad\textrm{with}\quad
 \beta_k \df \frac{|\vecp_{k+1}|}{p_{k+1}^0} 
       = \frac{\sqrt{\lambda(\Ss_{k+1},\sa_{k+1},\Ss_k)}}
              {\Ss_{k+1}-\sa_{k+1}-\Ss_k}
\;\;,
\label{defbeta}\end{equation}
then 
$\;p_{k+1}=\frac{1+\beta_k}{2}\,p^{(L)}_{k+1}
       + \frac{1-\beta_k}{2}\,p^{(S)}_{k+1}\;$,
and
\begin{equation}
\aot^{(k)}\df\frac{p_k\cdot p^{(L)}_{k+1}}{Q_k\cdot p^{(L)}_{k+1}}
          =\beta_k^{-1}\ao^{(k)}-\ho
 \quad\textrm{with}\quad
 \ho \df \frac{1-\beta_k}{2 \beta_k}\left(1+\frac{\sigma_k-s_{k-1}}{s_k}\right)
%  \ao^{(k)}=\beta_k\left(\aot^{(k)}+\frac{1-\beta_k}{2 \beta_k} 
% \left(1+\frac{\sigma_k-s_{k-1}}{s_k}\right)\right)\df\beta_k(\aot^{(k)}+\ho) 
\;\;.
\notag\end{equation}
% \begin{gather}
%  p_{k+1}=\left(q^0,\vecq\right)=\frac{1+\beta}{2} 
% \left(q^0,\vecq\frac{q^0}{|\vecq|}\right)  
% + \frac{1-\beta}{2} \left(q^0,-\vecq\frac{q^0}{|\vecq|}\right)
% \equiv \frac{1+\beta}{2} q_{1L}+ \frac{1-\beta}{2} q_{1S}
% \nl
% \tilde{a}_1=\frac{p_k\cdot q_{1L}}{Q_k\cdot q_{1L}}
% \;\;\;
% \ao^{(k)}=\beta\left(\tilde{a}_1+\frac{1-\beta}{2 \beta} 
% \left(1+\frac{\sigma_k-s_{k-1}}{s_k}\right)\right)
% \;\;\;
% \beta=\sqrt{\lambda\left(1,\frac{\sigma_{k+1}}{s_{k+1}},
% \frac{s_k}{s_{k+1}}\right)}
% \notag\end{gather}
Since the relation between $\ao^{(k)}$ and $\aot^{(k)}$ is linear, 
the two-body phase space is still expressible 
in terms of $\ao^{(k)}$ scaled by $\beta_k$.
The generic $\ao$-integral is given by
\begin{equation}
\int_{\ao^{\mathrm{min}}}^{\ao^{\mathrm{max}}}d\ao\, 
\frac{1}{(\ao+\ho)\sqrt{(\,\atp(\ao)+\hh(\ao)\,)(\,\atm(\ao)+\hh(\ao)\,)}} 
\;\;,
\label{a1g}\end{equation}
with $\atpm$ as defined in (\ref{defa2pm}), $\hh$ is a constant (for this
integral) or $\hh(\ao)=\ao$.
The {\em kinematical\/} integration limits, coming from the requirement that
$\atpm(\ao)$ are real, are given by
\begin{equation}
\aopm =\frac{1}{2}
\left( 1+\frac{\sa_k-\Ss_{k-1}}{\Ss_k}\pm 
\sqrt{\lambda\left(1,\frac{\sa_k}{\Ss_k},
                     \frac{\Ss_{k-1}}{\Ss_k}\right)}\;\right) \;\;.
\notag\end{equation}
In the massless case we get $\aom=0,
\aop=1-\Ss_{k-1}/\Ss_k$ and we have to impose a lower bound on $\ao$
given by $ \aoz=\sa_{k+1,k}/p_{k+1}\cdot Q_k$. 
In all cases for the form of $\hh(\ao)$, the $\ao$-integral is of the type
\begin{equation}
   \int_{\ao^{\mathrm{min}}}^{\ao^{\mathrm{max}}}d\ao\,
   \frac{1}{(\ao+\ho)\fao(\ao)} \quad\textrm{with}\quad
   \fao(\ao) \df \sqrt{\ao^2+2\vv\ao+\ww^2}
   \quad,\quad
   \vv^2<\ww^2
   \;\;,
\label{defa1int}\end{equation}
and the indefinite integral is given by
\begin{equation}
   \int d\ao\,\frac{1}{(\ao+\ho)\fao(\ao)}
   = \frac{1}{\fao(-\ho)}\,
     \log\left(\frac{\ao+\ho+\fao(\ao)-\fao(-\ho)}
                    {\ao+\ho+\fao(\ao)+\fao(-\ho)}\right)  \;\;,
\notag\end{equation}
which is analytically invertible. The definite integral (times $\pi/4$ from the 
$\at$-integral and a factor $1/\beta_k$ if $p_{k+1}$ is massive)
gives the
weight factors in the generation of $\ao,\at$ for the closed antenna.  
% 
% The generation
% is as follows:
% \begin{alignat}{1}
% \ao&=\frac{2 b f}{(f+a/b)^2-1}\;\;\;f=F_{-}^{1-\rho}F_{+}^{\rho}
% \nl
% F_{\mp}&=\left(\left(b^2-2a\, \aopm+(\aopm)^2\right)^{1/2}+\frac{b^2-2 a\, 
% \aopm}{b}
% \right)/\aopm
% \end{alignat}
% 
% where
% \begin{alignat}{2}
% & b=\left( b_0^2+\si^2 \frac{\sigma_k}{s_{k+1}}+2 \ho-\ho^2 \right)^{1/2} 
% \;\;
% b_0=\frac{1}{2}\left(1+\frac{s_{k-1}-\sigma_k}{s_{k+1}}+
% \left(1-\frac{s_{k-1}-\sigma_k}{s_{k+1}}\right) c\right) 
% &
% \nl
% &
% a=\frac{1}{2}\left(1-\frac{s_{k-1}-\sigma_k}{s_{k+1}}+
% \left(1+\frac{s_{k-1}-\sigma_k}{s_{k+1}}\right) c\right)+\ho 
% \;\;\;
% \ho=\frac{1-\beta}{2 \beta} \left(1+\frac{\sigma_k-s_{k-1}}{s_k}\right)
% \;\;\;
% &
% \notag\end{alignat}
% with a soft factor given by
% \[
% \mathcal{W}=\frac{\left(b^2-2a \ao+\ao^2\right)^{1/2}}{b\,\beta}
% \log(F_{+}/F_{-})
% \]
% 
% In the case of $s_{k-1}=0$ we have 
% \[
% \int_{a_{1}^{\mathrm{min}}}^{a_{1}^{\mathrm{max}}}  d\ao 
% \frac{1}{(\ao+\ho)\, \sqrt{(a^{+}_2+h) (a^{-}_2+h)}}
% \]
% where $h$ is chosen to regularize the subsequent $\at$ integration.
% The generation is now slightly modified with
% \[ 
% b=\left( b_0^2+2\,\left(\frac{1+c}{2}+c\,h\right)\, \ho+\ho^2+
% \left(\frac{1-c}{2}\right)^2\left(\frac{\sigma_k}{s_k}\right)^2
% +2\frac{\sigma_k}{s_k}\left(\frac{1+c}{2}-h+\ho\right)
% \left(\frac{1-c}{2}\right)
%  \right)^{1/2} 
% \]
% \[
% b_0=\frac{1+c}{2}+h
% \]
% \[
% a=\frac{1+c}{2}+h c +
% \left(\frac{\sigma_k}{s_k}\right)\left(\frac{1-c}{2}\right)+\ho
% \]
% 
%%%%%%%%%%%%%%%%%%%%%%%%%%%%%%%%%%%%%%%%%%%%%%%%%%%%%%%%%%%%%%%%%%%%%%%%
\countpar\paragraph{\parcount The {\boldmath $\ao$}-variable in the case 
                             {\boldmath $m=1$}}
%%%%%%%%%%%%%%%%%%%%%%%%%%%%%%%%%%%%%%%%%%%%%%%%%%%%%%%%%%%%%%%%%%%%%%%%
This refers to (\ref{1split}), where  we have $\co=-1$ and
$\at=\ao+\mu$ with   
\begin{equation}
\mu\df\frac{s_{k-1}-\sigma_k}{s_k}
\quad,\quad\textrm{so that}\quad
\int_{\ao^{\mathrm{min}}}^{\ao^{\mathrm{max}}}d\ao\, 
\frac{1}{\ao (1-\ao) (\ao+\mu)(1-\mu-\ao)}
\label{a1g1}\end{equation}
is the integral to be performed. 
Since the integrand is equal to 
\begin{equation}
   \frac{1}{\ao(\ao+\mu)}+\frac{1}{\ao(1-\mu-\ao)}
   +\frac{1}{(\ao+\mu)(1-\ao)}+\frac{1}{(1-\ao)(1-\mu-\ao)} \;\;,
\notag\end{equation}
we see that the generation of $\ao$ can be done easily using the multi-channel
technique with four channels with weight
\begin{equation}
w_i=\frac{1}{g^{(2)}_i e_i-g^{(1)}_i d_i}
\left( \log\left( \frac{g^{(1)}_i\ao^{\mathrm{max}}+d_i}
                       {g^{(2)}_i\ao^{\mathrm{max}}+e_i}\right)
      -\log\left( \frac{g^{(1)}_i\ao^{\mathrm{min}}+d_i}
                       {g^{(2)}_i\ao^{\mathrm{min}}+e_i}\right) \right) 
% \quad,\quad i=1,2,3,4
 \quad,
\notag\end{equation}
where
\begin{center}
\begin{tabular}{|c|c|c|c|c|}
\hline
$i$ & $d_i$ & $e_i$ & $g^{(1)}_i$ & $g^{(2)}_i$
\\
\hline
1 & 0 & $\mu$ & $+$ & $+$ 
\\
\hline
2 & 0 & $1-\mu$ & $+$ & $-$
\\
\hline
3 & $\mu$ & $1$ & $+$ & $-$
\\
\hline
4& $1$ & $1-\mu$ & $-$ & $-$
\\
\hline
\end{tabular}
\end{center}
% The generation proceeds through a multi-channel algorithm
% over four channels given generically by
% \begin{equation}
% \ao = g^{(1)}_i\frac{e_i f - d_i}{1-g^{(1)}_i g^{(2)}_i f}
% \;\;\;mbox{with}\;\;\;
% f=\left( \frac{g^{(1)}_i x_{+}+d_i}{g^{(2)}_i x_{+}+e_i}\right)^\rho
% \left( \frac{g^{(1)}_i x_{-}+d_i}{g^{(2)}_i x_{-}+e_i}\right)^{1-\rho}
% \label{multi}\end{equation}
% and the non-normalized {\it a priori \/}[D[D[[C[C[C[\/} weights
The soft factor will simply be equal to $\sum_{i=1}^{4}w_i$. The integration
over the azimuthal angle, replacing the $\at$-integration, gives an extra 
factor of $2\pi$.
\subsubsection{Antenna split}
%%%%%%%%%%%%%%%%%%%%%%%%%%%%%%%%%%%%%%%%%%%%%%%%%%%%%%%%%%%%%%%%%%%%%%%%
With $q_1\propto(1,0,01)$ and $q_2\propto(1,0,0,-1)$, the two-body phase space
integral (\ref{splitden}) assumes the form
\begin{equation}
\int d\Ss_1d\Ss_2\,dQ_z\,
    \frac{\Theta(\sqrt{\Ss}-\sqrt{\Ss_1}-\sqrt{\Ss_2}\,)\,
          \Theta(\,\Ss_1-\Ss_1^{(0)}\,)\,
          \Theta(\,\Ss_2-\Ss_2^{(0)}\,)}
         {\Ss_1\Ss_2\,
	  (\,E_1(\Ss_1,\Ss_2)^2-Q_z^2\,)(\,E_2(\Ss_1,\Ss_2)^2-Q_z^2\,)} \;\;,
\label{splitg}\end{equation}
% \begin{equation}
% \int ds_1\, ds_2\, dp_z\, \frac{1}{s_1 s_2}\;\; 
% \frac{1}{E_1^2-p_z^2}\;\;
% \frac{1}{E_2^2-p_z^2}\;\;
% \label{splitg}\end{equation}
with the energies
\begin{equation}
E_1(\Ss_1,\Ss_2)\df\frac{s+s_1-s_2}{2\sqrt{s}} \quad,\quad 
E_2(\Ss_1,\Ss_2)\df\sqrt{s}-E_1(\Ss_1,\Ss_2) \;\;,
\notag\end{equation}
and where $\Ss_{1,2}^{(0)}$ are the sums of the matrix elements $\sa_{ij}$
(\ref{sigma}) corresponding with the momenta in two antennas to be generated. 
$Q_z$ is integrated between the kinematical limits 
$\pm\sqrt{E_1^2-\Ss_1^2}$, and  
can be treated in a way similar to the one described in the previous paragraph, 
by multi-channeling over four channels with
\begin{center}
\begin{tabular}{|c|c|c|c|c|}
\hline
$i$ & $d_i$ & $e_i$ & $g^{(1)}_i$ & $g^{(2)}_i$
\\
\hline
1 & $E_1$ & $E_2$ &  $+$ & $+$ 
\\
\hline
2 & $E_1$ & $E_2$  &   $+$ & $-$
\\
\hline
3 & $E_1$ & $E_2$  &  $+$ & $-$
\\
\hline
4 & $E_1$ & $E_2$   & $-$ & $-$
\\
\hline
\end{tabular}
\end{center}
The final weight is the sum of the channel-weights, divided by the 
(dimensionful) factor $4E_1E_2$. 
The generation of $\Ss_1,\Ss_2$ is a specific case of a more general 
problem described in \Appendix{AppBs1s2}.

%%%%%%%%%%%%%%%%%%%%%%%%%%%%%%%%%%%%%%%%%%%%%%%%%%%%%%%%%%%%%%%%%%%%%%%%%%%%%%%
\subsection{Open antenna\label{AppB}}
%%%%%%%%%%%%%%%%%%%%%%%%%%%%%%%%%%%%%%%%%%%%%%%%%%%%%%%%%%%%%%%%%%%%%%%%%%%%%%%
The three variables needed to describe the open antenna are
$\Ss_{k-1}$, $\ao^{(k)}$ and  $\vhi^{(k)}$. In each CMF$_k$, the 
$\vhi^{(k)}$-variable should
be generated with uniform distribution between $0$ and $2\pi$, and the
$\ao^{(k)}$-variable should be distributed following $1/\ao^{(k)}$ between
\begin{align}
   a_{1,+}^{(k)} 
    &= \frac{ \Ss_k + \sa_k-\Ss_{k-1}
             +\beta_k\sqrt{\la(\Ss,\sa_k,\Ss_{k-1})}}{2\Ss_k} \notag\\ 
   a_{1,-}^{(k)} 
    &= \max\!\left[
       \frac{ \Ss_k + \sa_k-\Ss_{k-1}
             -\beta_k\sqrt{\la(\Ss,\sa_k,\Ss_{k-1})}}{2\Ss_k}
        \,,\,\frac{\sa_{k+1,k}}{p_{k+1}\cdot Q_k}\right] 
	\;\;,
\notag\end{align}
with $\beta_k$ as defined in (\ref{defbeta}). The normalized density for
these generations is equal to
\begin{equation}
   \frac{2\beta_k}{\pi\Ln_k\,a^{(k)}_1}
   = \frac{\sqrt{\lambda(\Ss_{k+1},\sa_{k+1},\Ss_{k})}}
          {\pi\Ln_k\,(p_{k+1}\cdot p_{k})}
   \quad\textrm{with}\quad 
   \Ln_k = \log(a_{1,+}^{(k)}/a_{1,-}^{(k)})
   \;\;.
\notag\end{equation}
This suggests to use
\begin{align}
   \gf_{k+1}(\Ss_{k}) 
   \df&\,\frac{\sqrt{\lambda(\Ss_{k+1},\sa_{k+1},\Sa_{k})}}
               {(\Ss_k-\Sa_k)\sqrt{\lambda(\Ss_{k+1},\sa_{k+1},\Ss_k)}} 
	       \notag\\
   =&\, \frac{d}{d\Ss_{k}}\,
      \log\!\left(-\frac{ \Ss_{k}-\Sa_{k}
                        +\sqrt{\lambda(\Ss_{k+1},\sa_{k+1},\Ss_{k})}
                        -\sqrt{\lambda(\Ss_{k+1},\sa_{k+1},\Sa_{k})}}
                       { \Ss_{k}-\Sa_{k}
                        +\sqrt{\lambda(\Ss_{k+1},\sa_{k+1},\Ss_{k})}
                        +\sqrt{\lambda(\Ss_{k+1},\sa_{k+1},\Sa_{k})}}\right)
\notag\end{align}
instead of (\ref{s1g}) for the case that $\beta_{k}\neq1$. 
The logarithm contributes to the weight as a soft factor again.
For small values of the squared masses $\sa_k$, the factor in the numerator 
will be canceled by 
$\Ss_{k+1}-\Sa_{k+1}$ in the denominator of $\gf_{k+2}(\Ss_{k+1})$.
Just as in the case of the closed antenna, we end up with one remaining, and
desirable, 
factor $\Ss_2-\Sa_2=2(p_2\cdot p_1)$ in the denominator of the 
open antenna density, which cannot be achieved by the generation of the 
$\ao^{(k)}$-variables.

Let us denote the soft factor coming from the $\Ss_k$-generation by 
\begin{equation}
   \Gf_{k+1} \df 
   \int_{\Lambda_{k}}^{\Ss_{k+1}-\Delta_{k}}d\Ss_k\,\gf_{k+1}(\Ss_{k})
   \;\;,
\notag\end{equation}
then the complete density resulting from the decomposition (\ref{openA}) is 
\begin{equation}
  \frac{d\Dd_0(Q_n;p_n,\ldots,p_1)}{d\Phi_{n}(Q_n;p_n,\ldots,p_1)} 
  = \frac{(q_1\cdot Q_n)\sqrt{\lambda(\Ss_n,\sa_n,\Sa_{n-1})}\;\Bb_{n}}
	 {(q_1\cdot p_{n})(p_{n}\cdot p_{n-1})\cdots
          (p_{3}\cdot p_{2})(p_{2}\cdot p_{1})} 
    \quad,
\notag\end{equation}
with soft factor
\begin{align}
   \Bb_{n} 
   \df \frac{1}
            {(4\pi)^{n-1}\Gf_{3}\Ln_{3}\Ln_{2}}
       \prod_{k=n}^{4}\frac{1}{\Gf_{k}\Ln_{k}}
       \sqrt{1-\frac{4\sigma_{k-1}\Sa_{k-2}}{(\Ss_{k-1}-\Sa_{k-1})^2}}
	 \;\;.
\notag\end{align}
The factor $p_1\cdot q_2$ is missing in the denominator of the density, which 
is the reason why we call this an {\em open\/} antenna. Other open antennas
can be obtained by starting with a decomposition of $Q_{n}$ into two momenta, 
from each of which open antennas of the above type are generated, with 
initial momentum $q_1$ for the one, and $q_2$ for the other. 

In order to digress about this procedure, let us extend the labeling a bit. 
With a set $\{I_1,I_2,\ldots,I_n\}$ of $n$ 
non-equal labels, we write
\begin{equation}
   Q_{I_k} \df \sum_{m=1}^kp_{I_m} \quad,\quad
   s_{I_k} \df Q_{I_k}^2           
\notag\end{equation}
and so on.  Now take $n=M+N$ and let 
\begin{equation}
  \{I_1,\ldots,I_{N}\} \df \{M+1,M+2,\ldots,M+N\} \;\;\;,\;\;\;
  \{J_1,\ldots,J_{M}\} \df \{M,M-1,\ldots,2,1\} \;.
\notag\end{equation}
We introduce $p_{I_{N+1}}=q_1$, $p_{J_{M+1}}=q_2$,  
and the decomposition
\begin{align}
  &ds_{J_{M}}ds_{I_{N}}\,
        d\Phi_2(Q_{n};\Ss_{I_{N}},\Ss_{J_{M}};Q_{I_{N}},Q_{J_{M}}) \notag\\
  &\times d\Dd_0(Q_{I_{N}};p_{I_{N}},p_{I_{N-1}},\ldots,p_{I_1})\,
          d\Dd_0(Q_{J_{M}};p_{J_{M}},p_{J_{M-1}},\ldots,p_{J_1})
\;\;,	 
\notag\end{align}
which produces the density
\begin{align}
   \frac{d\Dd_M(Q_n;p_n,\ldots,p_1)}{d\Phi_{n}(Q_n;p_n,\ldots,p_1)} 
   &=      \frac{2Q_n^2\;\Bb_{J_{M}}\Bb_{I_{N}}}
                {\pi\sqrt{\la(Q_n^2,s_{J_M},s_{I_N})}}
     \times\frac{(p_{M+1}\cdot p_{M})}
                {(q_1\cdot p_{n})(p_{n}\cdot p_{n-1})\cdots
                 (p_{2}\cdot p_{1})(p_{1}\cdot q_2)}\notag\\
   &\times(q_1\cdot Q_{I_{N}})(q_2\cdot Q_{J_{M}})
            \sqrt{\lambda(\Ss_{I_{N}},\sa_{I_{N}},\Sa_{I_{N-1}})
	          \lambda(\Ss_{J_{M}},\sa_{J_{M}},\Sa_{J_{M-1}})}
\;.
\label{defDN}\end{align}
In order to cancel the `undesirable' factors on the second line, we need to 
take care of the generation of 
$\Ss_{I_{N}},\Ss_{J_{M}},Q_{I_{N}},Q_{J_{M}}$.

%%%%%%%%%%%%%%%%%%%%%%%%%%%%%%%%%%%%%%%%%%%%%%%%%%%%%%%%%%%%%%%%%%%%%%%%%%%%%%%%
\subsubsection{Generation of {\boldmath $Q_{I_{N}},Q_{J_{M}}$}}
%%%%%%%%%%%%%%%%%%%%%%%%%%%%%%%%%%%%%%%%%%%%%%%%%%%%%%%%%%%%%%%%%%%%%%%%%%%%%%%%
Since $q_1\propto(1,0,0,1)$ and $q_2\propto(1,0,0,-1)$, we can write
\begin{equation}
   \ao = \frac{q_1\cdot Q_{I_{N}}}{q_1\cdot Q_n}
   \quad\textrm{and}\quad
   \frac{Q_{J_{M}}\cdot q_2}{Q_n\cdot q_2} = \ao + \mu
   \quad\textrm{with}\quad
   \mu = \frac{\Ss_{J_{M}}-\Ss_{I_{N}}}{\Ss_n}\;\;,
\notag\end{equation}
so that the generation of an azimuthal angle $\vhi$ between $0$ and $2\pi$ with
the uniform distribution, and the generation of $\ao$ with a density 
proportional to
\begin{equation}
   \frac{1}{\ao(\ao+\mu)}
   \quad\textrm{between}\quad
   \aopm = \frac{\Ss_n+\Ss_{I_{N}}-\Ss_{J_{M}}
                 \pm\sqrt{\lambda(\Ss_n,\Ss_{I_{N}},\Ss_{J_{M}})}}
                {2\Ss_n}
\notag\end{equation}
leads to the total density
\begin{equation}
   d\Phi_2(Q_n;\Ss_{I_{N}},\Ss_{J_{M}};Q_{I_{N}},Q_{J_{M}})\,
   \frac{2\mu}{\pi\log\frac{\aop(\aom+\mu)}{\aom(\aop+\mu)}}\,
   \frac{(q_1\cdot Q_n)(Q_n\cdot q_2)}
        {(q_1\cdot Q_{I_{N}})(Q_{J_{M}}\cdot q_2)}
   \;\;.
\notag\end{equation}

%%%%%%%%%%%%%%%%%%%%%%%%%%%%%%%%%%%%%%%%%%%%%%%%%%%%%%%%%%%%%%%%%%%%%%%%%%%%%%%%
\subsubsection{Generation of {\boldmath $\Ss_{I_{N}},\Ss_{J_{M}}$}}
%%%%%%%%%%%%%%%%%%%%%%%%%%%%%%%%%%%%%%%%%%%%%%%%%%%%%%%%%%%%%%%%%%%%%%%%%%%%%%%%
\countpar\paragraph{\parcount If {\boldmath $M=1$},} 
then $\Ss_{J_{M}}=\sa_{J_{M}}$, and we only need to generate 
$\Ss_{I_{N}}$ with a density proportional to 
\begin{equation}
   \frac{1}{\sqrt{\lambda(\Ss_{I_{N}},\sa_{I_{N}},\Sa_{I_{N-1}})}}
   = \frac{d}{d\Ss_{I_{N}}}
     \log\left(\Ss_{I_{N}}-\sa_{I_{N}}-\Sa_{I_{N-1}}
               +\sqrt{\lambda(\Ss_{I_{N}},\sa_{I_{N}},\Sa_{I_{N-1}})}\right)
   \;\;.
\notag\end{equation}
This density cancels the corresponding factor in the total antenna density. 
Something similar can be done if $N=1$. 

\newcommand{\cc}{c}
\newcommand{\mm}{m}
\newcommand{\Li}{\mathrm{Li}}
\countpar\paragraph{\parcount If {\boldmath $N>1$} and {\boldmath $M>1$},\label{AppBs1s2}} 
then both 
$\Ss_{I_{N}}$ and $\Ss_{J_{M}}$ have to be generated, in the region where 
$\sqrt{\Ss_{n}}-\sqrt{\Ss_{I_{N}}}-\sqrt{\Ss_{J_{M}}}>0$. 
This is far more complicated than the previous case, and we restrict ourselves
to a density with a denominator proportional to
$(\Ss_{I_{N}}-\Sa_{I_{N}})(\Ss_{J_{M}}-\Sa_{J_{M}})$. 
Because $\sa_{I_{N}}\Sa_{I_{N-1}}$ and $\sa_{J_{M}}\Sa_{J_{M-1}}$ may be 
considered small, it still cancels the factor 
\begin{equation}
   \sqrt{\lambda(\Ss_{I_{N}},\sa_{I_{N}},\Sa_{I_{N-1}})
              \lambda(\Ss_{J_{M}},\sa_{J_{M}},\Sa_{J_{M-1}})}
\notag\end{equation}
in (\ref{defDN}).
We shall write $\Ss_1,\Ss_2$ instead of $\Ss_{I_N},\Ss_{J_M}$ from now on, 
and denote $\mm\df\sqrt{\Ss_n}\,$, 
\begin{equation}
   \mm_{1}^2 \df  \Sa_{I_{N}}\;\;,\;\;\;
   \cc^2_{1} \df \sum_{K,L=1}^{N}\sa_{I_{K},I_{L}}
   \quad\textrm{and}\quad
   \mm_{2}^2 \df  \Sa_{J_{M}}\;\;,\;\;\;
   \cc^2_{2} \df \sum_{K,L=1}^{M}\sa_{J_{K},J_{L}}  \;\;.
\notag\end{equation}
We choose first to generate $\Ss_1$, and then $\Ss_2$, so that the integral,
corresponding with the generation, becomes
\begin{equation}
   \int_{\cc^2_1}^{(\mm-\mm_2)^2}\frac{d\Ss_1}{\Ss_1-\mm_1^2}
     \int_{\cc^2_2}^{(\mm-\sqrt{\Ss_1})^2}\frac{d\Ss_2}{\Ss_2-\mm_2^2} \;\;.
\notag\end{equation}
The $\Ss_2$-integral is simple, and shows that $\Ss_2$ can easily be obtained
by inversion. After integration over $\Ss_2$, the $\Ss_1$-integral becomes
\begin{equation}
   \int_{\cc^2_1}^{(\mm-\mm_2)^2}\frac{d\Ss_1}{\Ss_1-\mm_1^2}
     \left[ \log\left((\mm-\sqrt{\Ss_1})^2-\mm_2^2\right)
           -\log\left(\cc^2_2-\mm_2^2\right)\right]  \;\;.
\notag\end{equation}
The $\Ss_1$-variable distributed following this integrand can be obtained with 
high efficiency by rejection from 
the density proportional to $1/(\Ss_1-\mm_1^2)$. The total integral can be 
calculated, and is given by
\begin{multline}
  \sum_{\rho,\epsilon=\pm1}
      \left[\Li\left(\frac{x_1+\mm_1}{\mm+\rho\mm_2+\epsilon\mm_1}\right)
            +\log\left(\frac{\mm+\rho\mm_2-x_1}{\mm}\right)
             \log\left(\frac{x_1+\mm_1}{\mm+\rho\mm_2+\epsilon\mm_1}\right)
      \right]_{x_1=\cc_1}^{\mm-\mm_2} \\
  \hspace{12pt}- \log\left(\frac{\cc^2_2-\mm_2^2}{\mm^2}\right)
           \log\left(\frac{(\mm-\mm_2)^2-\mm_1^2}{\cc^2_1-\mm_1^2}\right)
	    \notag \;\;,
\end{multline}
where 
\begin{equation}
   \Li(x) \df \sum_{n=1}^\infty\frac{(1-x)^n}{n^2} \quad.
\notag\end{equation}

\subsection{SARGE\label{AppSARGE}}
\newcommand{\ipb}[2]{(#1\cdot#2)}
\newcommand{\Bo}{\mathcal{H}}
\newcommand{\Ro}{\mathcal{R}}
\newcommand{\Ord}{\mathcal{O}}
\newcommand{\Pol}{\mathbf{P}}
\newcommand{\ind}{\vartheta}
\newcommand{\Mod}[1]{\lceil #1\rceil}
\newcommand{\nout}{n}
\newcommand{\ximax}{\xi_{\mathrm{m}}}
\newcommand{\scm}{s}
\newcommand{\Bnt}{B}
\newcommand{\scut}{s_0}
In this Appendix, we give a short review about {\tt SARGE} and present the 
adaptations applied in {\tt SARGE.n}. We need to start with the establishment of 
some notation.

$\Bo_p$ is a Lorentz transformation that boosts momentum $p$ to 
$(\sqrt{p^2},0,0,0)$.
$\Ro_p$ is a Lorentz transformation that rotates momentum $p$ to 
$(p^0,0,0,|\vecp|)$. 
The standard representation of a unit vector in terms of parameters 
$z\in[-1,1]$ and $\vhi\in[0,2\pi]$ is denoted by
\begin{equation}
   \Hat{n}(z,\vhi)\df(\sqrt{1-z^2}\,\sin\vhi,\sqrt{1-z^2}\,\cos\vhi,z) \;\;.
\notag\end{equation} 
$\Pol_n\subset[-1,1]^n$ is the $n$-dimensional polytope, which consists of the 
support of the indicator function
\begin{equation}
\ind_{\Pol_n}(x_1,x_2,\ldots,x_n)
   \df\prod_{i=1}^n\Theta(1-|x_i|)\prod_{j,k=1}^{n}\Theta(1-|x_j-x_k|) \;\;.
\notag\end{equation} 
The algorithm for the generation of $\nout$ final-state momenta with a
single antenna structure without initial-state momenta as given in~\cite{sarge} 
is
\begin{Alg}
\begin{enumerate}
\item generate two massless momenta $q_1,q_{\nout}$, back-to-back;
\item generate $(x_1,\ldots,x_{2\nout-4})\in\Pol_{2\nout-4}$\; 
      and $(\vhi_2,\ldots,\vhi_{\nout-1})\in[0,2\pi]^{\nout-2}$, all 
      uniformly distributed;
\item for $i=2,\ldots,\nout-2$ construct $q_i$ following 
      (with $x_0\df0$) 
      \begin{gather}
      b_1\lar q_{i-1}+q_{\nout}
      \quad,\quad
      b_2\lar \Bo_{b_1}q_{i-1}
      \quad,\notag\\
      \xi_1\lar e^{(x_{2i-3}-x_{2i-4})\log\ximax}
      \quad,\quad
      \xi_2\lar e^{(x_{2i-2}-x_{2i-4})\log\ximax}
      \quad,\notag\\
      v^0\lar\sqrt{\frac{\ipb{q_{i-1}}{q_{\nout}}}{2}}\,(\xi_2+\xi_1)
      \quad,\quad
      z\lar\frac{\xi_2-\xi_1}{\xi_2+\xi_1}\quad,\quad
      \vecv\lar q^0\Ro_{b_2}^{-1}\Hat{n}(z,\vhi_i)
      \quad,\notag\\
      q_i\lar\Bo_{b_1}^{-1}v\quad;
      \notag\end{gather}
\item put $p_i \lar u\Bo_{Q}q_i$ for $i=1,\ldots,\nout$, where
      $Q=\sum_{i=1}^{\nout}q_i$ and $u=\sqrt{\scm/Q^2}$\;.
\end{enumerate}
\label{Alg00}
\end{Alg}
The density with which the generated momenta are distributed is then given by 
\begin{equation}
   d\Phi_{\nout}(P;p_1,\ldots,p_\nout)\,
   \frac{\scm^2}{2\pi^{\nout-2}}\,
   g_{\nout}(p_1,p_2,\ldots,p_{\nout})\Ant_{\nout}(p_1,p_2,\ldots,p_{\nout}) 
   \quad,
\label{EqnAnt00}   
\end{equation}
with $P=(\sqrt{\scm},0,0,0)$, where 
$\Ant_{\nout}(p_1,p_2,\ldots,p_{\nout})$ is defined in (\ref{defAnt}),
 and where
% \begin{equation}
%    \Ant_{\nout}(p_1,p_2,\ldots,p_{\nout})
%    \df \frac{1}{\ipb{p_{\nout}}{p_1}\ipb{p_1}{p_2}\ipb{p_2}{p_3}\cdots
%                 \ipb{p_{\nout-1}}{p_{\nout}}}
% \quad,
% \notag\end{equation}
% and
\begin{equation}
  g_{\nout}(p_1,p_2,\ldots,p_{\nout})
  \df\frac{\ind_{\Pol_{2\nout-4}}(x_1,x_2,\ldots,x_{2\nout-4})}
          {(2\nout-3)(\log\ximax)^{2\nout-4}}
  \quad,
\notag\end{equation}
with   
\begin{equation}
  x_{2i-3} = \frac{1}{\log\ximax}\log\frac{\ipb{p_{i-1}}{p_{i}}}
                     {\ipb{p_1}{p_{\nout}}}
  \quad\textrm{and}\quad
  x_{2i-4} = \frac{1}{\log\ximax}\log\frac{\ipb{p_{i}}{p_{\nout}}}
                     {\ipb{p_1}{p_{\nout}}}
 \;\;,\quad i=2,\ldots,\nout-2
 \;\;.
\notag\end{equation}
The density obtained before step 4 of \Algorithm{Alg00} is invariant under
simultaneous scaling of all momenta, but the sum of the momenta is not at rest.
In order to achieve this, the scaling symmetry has to be broken. In order to
include the initial-state momenta in the density, more symmetries have to be
broken: the symmetry under cyclic permutations of the momenta, a part of
the simultaneous rotation symmetry of the momenta, and finally the symmetry 
under the reflection permutation 
$(1,2,\ldots,\nout)\mapsto(\nout,\nout-1,\ldots,1)$.

Let us denote $\Mod{k}=k\!\!\mod\nout$. The cyclic symmetry is broken through
\begin{Alg}
\begin{enumerate}
\item choose a $k\in\{0,2,\ldots,\nout-1\}$ with relative probability
      $\ipb{p_{\Mod{k}}}{p_{k+1}}$ \;;
\item put $\{p_{1},p_{2},\ldots,p_{\nout}\}
           \lar\{p_{\Mod{1+k}},p_{\Mod{2+k}},
	         \ldots,p_{\Mod{\nout+k}}\}$\;.
\end{enumerate}
\label{Alg01}
\end{Alg}
As a result, the function $\Ant_{\nout}$ in the density (\ref{EqnAnt00}) 
is replaced by 
\begin{equation}
   \frac{\nout}{\ipb{p_{\nout}}{p_{1}}+\ipb{p_{1}}{p_{2}}
               +\cdots+\ipb{p_{\nout-1}}{p_{\nout}}}
      \times\ipb{p_\nout}{p_1}\Ant_{\nout}(p_1,\ldots,p_{\nout})
 \;\;.
\notag\end{equation}
The factor $\ipb{p_{\nout}}{p_{1}}$ in the denominator of $\Ant_{\nout}$ is 
replaced by the average of all factors. Next, part of the rotation 
symmetry and the symmetry under the reflection permutation are broken through
\begin{Alg}
\begin{enumerate}
\item generate $\vhi\in[0,2\pi]$ uniformly distributed, and $z\in[-1,1-c]$ 
      following $1/(1-z)$\;;
\item rotate all momenta such that $\vecp_1$ lies along $\Hat{n}(z,\vhi)$\;.
\end{enumerate}
\label{Alg02}
\end{Alg}
If we denote $q_1\df\srac{1}{2}\sqrt{\scm}(1,0,0,1)$, this leads to an extra
factor 
\begin{equation}
    \frac{1}{2\pi\log(2/c)}\times
    \frac{\srac{1}{2}\sqrt{\scm}\,p_1^0}{\ipb{q_1}{p_{1}}}
\notag\end{equation}
in the density. The cut-off $c$  can be taken equal to 
$\scut/(\sqrt{\scm}\,p_1^0)$, where $\scut$ should be a cut-off on the 
invariant mass $(q_1+p_1)^2$. Notice that $p_1^0=\ipb{P\,}{p_1}/\sqrt{\scm}$,
so that the new factor in the density becomes
\begin{equation}
   \frac{1}{4\pi\log(\,2\ipb{P\,}{p_1}/\scut\,)}
   \times\frac{\ipb{P\,}{p_1}}{\ipb{q_1}{p_{1}}} 
   \quad.
\notag\end{equation}
In order to include a factor proportional
to $\ipb{p_{\nout}}{q_2}$ in the denominator of the density, where 
$q_2\df\srac{1}{2}\sqrt{\scm}(1,0,0,-1)$, \Algorithm{Alg02} can be 
preceded by
\begin{Alg}
\begin{enumerate}
\item choose with equal probabilities whether to rotate all momenta 
      such that $\vecp_1$ lies along $\Hat{n}(z,\vhi)$, 
      or that $\vecp_{\nout}$ lies along $-\Hat{n}(z,\vhi)$.
\end{enumerate}
\label{Alg03}
\end{Alg}
The total density (\ref{EqnAnt00}) becomes then such that 
$\Ant_{\nout}$ is replaced by 
\begin{equation}
  \frac{\nout}{4\pi}\times
  \frac{ \frac{\ipb{P\,}{p_1}\ipb{p_{\nout}}{q_2}}
              {\log(\,2\ipb{P\,}{p_1}/\scut\,)}
        +\frac{\ipb{q_1}{p_{1}}\ipb{P\,}{p_{\nout}}}
	      {\log(\,2\ipb{P\,}{p_{\nout}}/\scut\,)}}
             {\ipb{p_{\nout}}{p_{1}}+\ipb{p_{1}}{p_{2}}
             +\cdots+\ipb{p_{\nout-1}}{p_{\nout}}}
  \times\Ant_{\nout+2}(q_1,\;p_1,p_2,\ldots,p_{\nout-1},p_{\nout},\;q_2)
\;\;.
\notag\end{equation}
In the end, we want to obtain all permutations in the momenta
$\{p_1,p_2,\ldots,p_{\nout},q_2\}$ of the above density. Some of them can be
obtained by relabeling, and for the others, which change the position of $q_2$,
we can do the following. If the cyclic permutation in \Algorithm{Alg01} is chosen
with relative probability
$\ipb{p_{\Mod{k}}}{p_{k+1}}\ipb{p_{k+i}}{p_{k+1+i}}$, then the factor 
$\ipb{p_{\nout}}{p_{1}}\ipb{p_{i}}{p_{i+1}}$ in the 
denominator of $\Ant_{\nout}$ is replaced by the average over its cyclic 
permutations. Then we can choose with equal relative probabilities whether 
to rotate all final-state momenta such that 
$\vecp_1$ lies along $\Hat{n}$, 
$\vecp_{i}$ lies along $-\Hat{n}$, 
$\vecp_{i+1}$ lies along $-\Hat{n}$, or 
$\vecp_{\nout}$ lies along $\Hat{n}$. This will lead to density that is 
proportional to
\begin{equation}
  \Ant_{\nout+2}( q_1,\;p_1,p_2,\ldots,p_{i-1},p_i,\;q_2,\;p_{i+1},p_{i+2},
                 \ldots,p_{\nout-1},p_{\nout})
\;\;.
\notag\end{equation}

\subsection{Permutations}
For the analyses of the the multi-channel procedure over the antenna densities
for the different permutations of the momenta, we need an enumeration of the 
permutations. In other words, we need a mapping 
\begin{equation}
   \{1,2,\ldots,n!\}\mapsto \mathit{Sym}_n \;\;.
\notag\end{equation}
During a computation, one could, of course, just store an enumeration in the
memory of the computer, but this costs an amount of memory of $\Ord(n!)$.
Every algorithm that delivers all permutations supplies a mapping 
{\it a priori\/}, but this algorithm has an {\it a priori\/} computational 
complexity of $\Ord(n!)$. 

We propose the following algorithm as a solution.
Any number $l\in\{1,2,\ldots,n!\}$ can uniquely be written in the basis of the
lower factorials: $l=1+l_2+l_32!+l_43!+\cdots+l_{n}(n-1)!$, where
$l_k\in\{0,1,\ldots,k-1\}$. 
Let $\gamma^{(k)}_i$ denote the $i$-fold cyclic permutation of the first $k$
elements of $(1,2,\ldots,n)$, for example $\gamma^{(3)}_2(1,2,3,4)=(2,3,1,4)$.
Then any
permutation of $(1,2,\ldots,n)$ can be written as
$\gamma^{(2)}_{l_2}\gamma^{(3)}_{l_3}\cdots\gamma^{(n)}_{l_{n}}(1,2,\ldots,n)$. 
This leads to a  mapping of the required kind, which has a complexity of
$\Ord(n^2)$. In a loop delivering all permutations, it can easily be reduced to
complexity $\Ord(n)$ at the cost of an amount of memory of $\Ord(n)$.

\end{document}